\documentclass[11pt]{article}
\usepackage{amsmath}
\usepackage{graphicx,psfrag,epsf}
\usepackage[utf8]{inputenc}
\usepackage{enumerate}
\usepackage[dcucite]{harvard}  

\newcommand{\blind}{0}

\newtheorem{theorem}{Theorem}
\newtheorem{lemma}{Lemma}

\newtheorem{corollary}{Corollary}

\newcommand{\be}{\begin{equation}} 
\newcommand{\ee}{\end{equation}} 
\newcommand{\E}{\mbox{E}} 
\newcommand{\Var}{\mbox{Var}} 
\newcommand{\Cov}{\mbox{Cov}}
 
\newcommand{\bq}{\begin{equation}} 
\newcommand{\eq}{\end{equation}}

\newcommand{\bc}{\begin{center}} 
\newcommand{\ec}{\end{center}} 
\newcommand{\bi}{\begin{itemize}} 
\newcommand{\ei}{\end{itemize}} 
\newcommand{\VEKK}[1]{} 

\newcommand{\cv}{\gamma}
\newcommand{\vekk}[1]{} 
\newcommand{\A}{A}  
\newcommand{\Y}{\tau - T_{N(\tau)}}

\let\citeN=\citeasnoun

\begin{document}

\def\spacingset#1{\renewcommand{\baselinestretch}%
{#1}\small\normalsize} \spacingset{1}


\if0\blind
{
  \title{ A Class of Tests for Trend in Time Censored Recurrent
    Event Data}
  \author{Jan Terje Kval\o y \\ Department of Mathematics and Physics \\
University of Stavanger, Norway \\  \\ 
Bo Henry Lindqvist \\
Department of Mathematical Sciences \\ Norwegian University of Science
and Technology, Norway \\ }
\date{}
  \maketitle
} \fi

\if1\blind
{
  \bigskip
  \bigskip
  \bigskip
  \begin{center}
    {\LARGE\bf A Class of Tests for Trend in Time Censored\\[5mm] Recurrent Event Data}
\end{center}
  \medskip
} \fi

\bigskip
\begin{abstract}
  Statistical tests for trend in recurrent event data not following a
  Poisson process are generally constructed for event censored
  data. However, time censored data are more frequently encountered in
  practice. In this paper we contribute to filling an important gap in
  the literature on trend testing by presenting a class of statistical
  tests for trend in time censored recurrent event data, based on the
  null hypothesis of a renewal process.  The class of tests is
  constructed by an adaption of a functional central limit theorem for
  renewal processes. By this approach a number of tests for time
  censored recurrent event data can be constructed, including among
  others a  version of the classical Lewis-Robinson trend test and an
  Anderson-Darling type test. The latter test turns out to have
  attractive properties for general use by having good power
  properties against both monotonic and non-monotonic
  trends. Extensions to situations with several processes are
  considered. Properties of the tests are studied by simulations, and
  the approach is illustrated in two data examples.
\end{abstract}

\noindent%

   
\section{Introduction}
\label{sec:intro}

Analyzing recurrent event data is a challenge encountered in many
fields, for instance engineering, medicine and economy to mention
some. Generally, recurrent event data arise when the phenomenon
studied can occur repeatedly. Some examples are the occurrence of a
failure in a repairable system or the outbreak of a recurrent
disease. One aspect of the data which typically is of interest is to
examine whether there are any systematic alterations, i.e., trends, in
the pattern of events. For example, does a repairable system have a
tendency to fail more often as it gets older? Or is there any
improvement in how often a recurrent disease occurs for a particular
patient? Visual inspections of the data can be very useful and give
important information on systematic tendencies in the data, but
generally, in order to distinguish actual systematic alterations from
random fluctuations, statistical methods are needed.

There is a rich literature on trend testing, see for instance the
overviews in \citeN{coxlewis}, \citeN{ascherfeingold},
\citeN{kvaloeylindqvist} and \citeN{Lawless2012tmt}. Trend tests are
based on different assumptions for the data collection process and
different definitions of trend.  Many of the existing tests for trend
are based on Poisson process theory and constructed for testing the
null hypothesis of a homogeneous Poisson process (HPP), see for
instance \citeN{coxlewis}, \citeN{ascherfeingold},
\citeN{cohensackrowitz}, \citeN{kvaloeylindqvist},
\citeN{Lawless2012tmt} and references therein. Such tests are,
however, generally sensitive to departures from the Poisson process
assumption.  This fact was noted in the classical reference
\citeN{lewisrobinson}, who observed that the commonly used Laplace
trend test often led to rejection of the null hypothesis of no trend,
even in cases where a trend could not exist. More specifically, the
authors observed that false rejections were particularly occurring in
cases of overdispersion of the interevent times with respect to the
exponential distribution. Their idea was to modify the Laplace test
statistic to account for this overdispersion, which led to the test
known under the name of Lewis-Robinson test, to be further considered
later in this paper. 
  
 The immediate conclusion to draw from this seems to be that, unless
 the Poisson assumption can be verified, trend tests need to be based
 on more general null hypotheses than the one of HPP. So how could one
 formalize a more useful null hypothesis?  \citeN{Lawless2012tmt}
 concluded that there is no single definition which covers all cases
 that can naturally be thought of. \citeN{lewisrobinson} argued that a
 definition of no trend should state that the event process is
 stationary in some sense, possibly allowing some amount of serial
 correlation.  On the other hand, because of analytical possibilities
 they found that the renewal process (RP) assumption would be the best
 choice for further investigations. Under this assumption they were
 able to repair the Laplace test and introduce the Lewis-Robinson test.

 In this paper we shall consider trend tests assuming the null
 hypothesis of RP. In addition to the Lewis-Robinson test, there exist several
 trend tests in the literature based on this null hypothesis. We would
 like to mention first the nonparametric test by \citeN{mann}.  Other
 tests are found in \citeN{ascherfeingold}, \citeN{klRP},
 \citeN{vaurio2}, \citeN{Lawless2012tmt} and references therein.
  
 RP based tests for trend, including the classical Lewis-Robinson test
 are, however, usually constructed for \textit{event censored} data,
 which means that the recurrent event process is censored when it has
 completed a fixed number of renewal events. On the other hand,
 \textit{time censored} data, where the event process is censored
 after a predetermined observation period, are far more naturally
 occurring in practice. As pointed out by \citeN{Lawless2012tmt},
 there is still an unfortunate lack of available trend tests
 constructed for time censored data. The crucial issue when going from
 event censoring to time censoring is how to involve in a consistent
 manner the time interval from the last event to the censoring
 time. \citeN{Lawless2012tmt} argued that ignoring this interval may
 lead to considerable bias, see also the most interesting discussion
 of this and related issues in \citeN{aalenhusebye}. The latter
 authors, furthermore, pointed out that it is far less critical to
 ignore an incomplete time at the start of the observation, which will
 not introduce bias although it might incur a certain loss of
 efficiency.

 With the above as our motivation and point of departure, we
 demonstrate in this paper how a flexible class of trend tests for
 time censored data can be constructed under the RP null
 hypothesis. We thereby complement the above mentioned literature on
 trend tests for event censored data, in particular the paper by
 \citeN{Lawless2012tmt}. 
 Our construction is based on an adaption of Donsker's theorem
 \cite{Donsker1952} to renewal processes following the lines of
 \citeN{billingsley1999convergence}. Among other tests, the class
 turns out to include a time censored version of the Lewis-Robinson
 test, an Anderson-Darling type test with power against both monotonic
 and non-monotonic trends and an extension of the Lewis-Robinson test
 with power against non-monotonic trend.  After having studied tests for trend
 in single processes, we consider extensions to trend tests based on
 the joint observation of several processes.

The paper is organized as follows. In Section~\ref{sec:donsker} we
define the necessary notation and give some key results for renewal
processes. The general construction of tests is presented in
Section~\ref{sec:classof} and  several specific tests are derived. 
 Section~\ref{sec:several} discusses extensions to cases where several similar
processes are observed.  A simulation study is presented in Section~\ref{sec:sim}, while two case studies are considered in Section~\ref{sec:case}. Some concluding remarks are given in Section~\ref{sec:conc}. The paper is ended by Appendix 1 and 2 providing detailed derivations of, respectively, parameter estimators and a specific trend test. 



\section{The Basic Convergence Results for Renewal Processes}
\label{sec:donsker}

\subsection{Setup and Notation}
\label{sec:notation}

Consider a renewal process observed from time $t=0$.  The
successive event times are denoted $T_1,T_2,\ldots,$ and the
corresponding interevent times, or gap times, are denoted
$X_1,X_2,\ldots$ where $X_i=T_i-T_{i-1},\; i=1,2,\ldots$ (with the convention $T_0=0$). The $X_i$ are independent and identically distributed, with $\E(X_i)=\mu$ and $\Var(X_i)=\sigma^2$, where it will be assumed throughout the paper that $\sigma^2 < \infty.$

We use the standard notation where $N(t)$ is the number of events in $(0,t]$ for all $t>0$. For the theory of renewal processes we refer to, e.g.,  \citeN{ross}  and \citeN{gallager}.

\subsection{A Functional Central Limit Theorem for Renewal Processes}
\label{subsec:timetrunc} 

The key result in our approach is a functional central limit theorem given in \citeN{billingsley1999convergence}. With notation as above, define
\[
V_{t,\mu,\sigma}(s)=\mu^{3/2}\frac{N(st)-st/\mu}{\sigma\sqrt{t}} \mbox{ for } 0\leq s\leq 1, \; t>0.
\]
 Then \cite[thm. 14.6]{billingsley1999convergence}, 
\begin{equation}
\label{first} 
V_{t,\mu,\sigma}\Rightarrow W  \mbox{ as } t\to\infty,
\end{equation}
  where
$\Rightarrow$ denotes weak convergence
and $W$ is the Wiener measure  \cite[chap. 8]{billingsley1999convergence}.

Now define $W^0(s)=W(s)-sW(1)$ for $0 \le s \le 1$, so that $W^0$ is a Brownian
bridge \cite[chap. 8]{billingsley1999convergence}. It is straightforward to verify that (\ref{first}) implies the following result which together with the succeeding corollary is the basis of our construction of trend tests.
 \begin{theorem}
\label{thm1}
Define
\begin{equation}
V_{t,\mu,\sigma}^0(s) = V_{t,\mu,\sigma}(s)-sV_{t,\mu,\sigma}(1)
=\mu^{3/2}\frac{N(st)-sN(t)}{\sigma\sqrt{t}} \mbox{ for $0 \le s \le 1$.}
\label{eq:bbtime}
\end{equation}
 Then $V_{t,\mu,\sigma}^0\Rightarrow W^0$.
\end{theorem}

Let the coefficient of variation of the interevent times $X_i$ be denoted  $\cv \equiv \sigma/\mu$. As will become clear, $\cv$ plays a special role in our construction of tests. First, define 
\begin{equation}
\label{tildeeq}
      \tilde V_{t,\cv}^0(s) =\frac{1}{\cv} \frac{N(st)-sN(t)}{\sqrt{N(t)}}  \mbox{ for $0 \le s \le 1$.}
\end{equation}
Then Theorem~\ref{thm1} implies the following corollary:

\begin{corollary}
\label{cor1}
With notation as above we have
\[
\tilde V_{t,\cv}^0 \Rightarrow W^0 \mbox{ as } t \rightarrow \infty.
\]
\end{corollary}

\noindent
\textit{Proof: }
We can write
\[
    \tilde V_{t,\cv}^0(s) = \frac{\sqrt{1/\mu}}{\sqrt{N(t)/t}}   V_{t,\mu,\sigma}^0(s).
\]
From standard renewal process theory \cite{ross} it is well known that $N(t)/t \rightarrow 1/\mu$ a.s. The result then follows by use of \citeN[thm. 3.1]{billingsley1999convergence}, sometimes called 'the converging together lemma'. The argument, using the uniform norm, is as follows:
\[
  \sup_{0 \le s \le 1} |\tilde V_{t,\cv}^0(s) - V_{t,\mu,\sigma}^0(s)| 
\le \left| \frac{\sqrt{1/\mu}}{\sqrt{N(t)/t}}-1 \right|  \sup_{0 \le s \le 1} |V_{t,\mu,\sigma}^0(s)| \stackrel{p}{\rightarrow} 0,
\] 
where the convergence to 0 follows since the first factor tends to 0 a.s. and hence in probability, and the last factor converges in distribution to $ \sup_{0 \le s \le 1} |W^0(s)|$ which has the Kolmogorov distribution (and will be considered below).

\section{The Class of Tests for Trend}
\label{sec:classof}

In the present section we consider event data from a single counting
process $N(t)$ observed from time $t=0$ until time censoring at the
given time $\tau > 0$. With notation as in Section~\ref{sec:donsker},
we thus observe a random number $N(\tau)$ of events, at times
$T_1,T_2,\ldots,T_{N(\tau)}$, and with fully observed interevent times
$X_1,X_2,\ldots,X_{N(\tau)}$ and a censored interevent time $\tau -
T_{N(\tau)}$.

From Theorem~\ref{thm1} and Corollary~\ref{cor1} it follows that, under
the null hypothesis of RP, $V_{\tau,\mu,\sigma}^0$ and $\tilde V_{\tau,\cv}^0$ will approximately be Brownian
bridges. Thus, if there is a trend in the data,
these processes are likely to deviate from a Brownian
bridge. Tests for trend can therefore be based on measures of deviation
from a Brownian bridge of the two asymptotically equivalent processes $V_{\tau,\mu,\sigma}^0$ and $\tilde V_{\tau,\cv}^0$. 

Since the parameters $\mu, \sigma, \cv$ are generally unknown, they
must be estimated. It is clear that the results of Theorem~\ref{thm1}
and Corollary~\ref{cor1} continue to hold under the RP assumption if
$\mu$, $\sigma$ and $\cv$ are replaced by consistent estimators,
$\hat\mu$,$\hat{\sigma}$ and $\hat \cv$.

Below we first  derive test statistics based on four different ways of
measuring deviations from a Brownian bridge.  This leads to test
statistics of, respectively, Lewis-Robinson, Kolmogorov-Smirnov,
Cram$\acute{\mbox{e}}$r-von Mises and Anderson-Darling types. In
addition we propose an extension of the Lewis-Robinson test which
can be used to construct tests for non-monotonic trend.  The test
constructions are based on applications of
Corollary~\ref{cor1}. Finally we discuss how to estimate the
parameters $\mu, \sigma$ and $\cv$.

\subsection{Lewis-Robinson Type Test}
\label{subsec:lr}

A classical measure of deviation from a Brownian bridge is the signed
area under the path of the process. Using Corollary~\ref{cor1} this gives rise to the statistic $
\int_0^1 \tilde V_{\tau,\hat\cv}^0(s)ds $, which  converges in distribution to  
$\int_0^1 W^0(s)ds $, which is normally distributed with expectation 0
and variance 1/12. 

In order to obtain the test statistic on the form that is most common for this test, we use instead the negative of the above suggested statistic, which  will have the same limiting distribution.
 By scaling we obtain an asymptotically standard
normally distributed test statistic given by
\begin{equation}
\label{onestar}
LR = -\sqrt{12}\int_0^1 \tilde V_{\tau,\hat\cv}^0(s)ds
= \frac{1}{\hat \cv} \cdot \frac{\sqrt{12}}{\tau \sqrt{N(\tau)}}
 \left[\sum_{i=1}^{N(\tau)} T_i - \frac{N(\tau)}{2}\tau
\right].
\end{equation}
If the factor $1/\hat \cv$ is ignored, we actually get the well known Laplace
test statistic for the null hypothesis of HPP for the time censored
case, which can be derived from properties of Poisson-processes. The
division by $\hat \cv$ corresponds to the correction obtained by \citeN{lewisrobinson}, who considered the event censored case. 
 
The resulting test will primarily have power
against deviations from an RP caused by monotonic
trends. It is seen that positive (negative) values of the test statistic will correspond to an increasing (decreasing) trend.

\subsection{Kolmogorov-Smirnov Type Test}

Another classical measure of deviation from a Brownian bridge is the
maximum deviation, giving rise to the statistic $
\sup_{s\in [0,1]} |\tilde V_{\tau,\hat\cv}^0(s)| $. By Corollary~\ref{cor1},
this statistic converges in distribution to
$\sup_{s\in [0,1]} |W^0(s)| $, which has the Kolmogorov distribution
\cite{kolmogorov,smirnov}. 
A Kolmogorov-Smirnov type test for trend in the time censored case is hence given by the test statistic
\begin{eqnarray}
KS  &=&\sup_{s\in [0,1]} |\tilde V_{\tau,\hat\cv}^0(s)|
=\frac{1}{\hat \cv} \frac{1}{\sqrt{N(\tau)}}
  \sup_{s\in [0,1]} |N(s\tau)-sN(\tau) |  \label{KSsup}   \nonumber\\
  &=& \frac{1}{\hat \cv} \frac{1}{\sqrt{N(\tau)}}
  \max_{i=1,\ldots,N(\tau)} \left\{ \max \left[ \; \; \left| i - \frac{N(\tau)}{\tau}T_i \right|, \left| i-1 - \frac{N(\tau)}{\tau}T_i\right| \; \;  \right] \right\}  .
\end{eqnarray}

\subsection{Cram$\acute{\mbox{e}}$r-von Mises Type Test}

Using the Cram$\acute{\mbox{e}}$r-von
Mises type measure we obtain 
$$ CvM  =
\int_0^1 \tilde V_{\tau,\hat\cv}^0(s)^2ds
\stackrel{d}{\rightarrow} \int_0^1 W^0(s)^2ds, $$ 
where the right hand side has the commonly known
limit distribution of the Cram$\acute{\mbox{e}}$r-von Mises statistic
\cite{andersondarling}.  Due to the squaring of
$\tilde V_{\tau,\hat\cv}^0(s)$ it is clear that a test which
rejects the null hypothesis of RP for large values of
$CvM$ will have sensitivity against both monotonic and non-monotonic
trends. Straightforward calculations give the statistic
\begin{equation}
CvM
= \frac{1}{\hat \cv^2}\frac{1}{N(\tau)}
\left\{\sum_{i=0}^{N(\tau)-1}\left[i^2\frac{X_{i+1}}{\tau}-i
      N(\tau)\frac{T_{i+1}^2-T_{i}^2}{\tau} \right]
+N(\tau)^2\left[\frac{T_{N(\tau)}^2}{\tau^2}-\frac{T_{N(\tau)}}{\tau}+\frac{1}{3} \right]
\right\} .
\label{eq:CV}
\end{equation}

\subsection{Anderson-Darling Type Test}

The Anderson-Darling type measure leads to
$$ AD  =
\int_0^1 \frac{V_{\tau,\hat\cv}^0(s)^2}{s(1-s)}ds
\stackrel{d}{\rightarrow} \int_0^1 \frac{W^0(s)^2}{s(1-s)}ds, $$ 
which has the limit distribution of the Anderson-Darling
statistic \cite{andersondarling,andersondarling2}. \
As for the Cram$\acute{\mbox{e}}$r-von Mises type test it is clear
that this test will have sensitivity against both monotonic and non-monotonic
trends. The difference between the Cram$\acute{\mbox{e}}$r-von Mises
and the Anderson-Darling statistics is that the latter
 puts more weight on
the information at the beginning and the end of the observation
interval. Straightforward but somewhat tedious calculations give that
\begin{eqnarray}
AD &=& \frac{1}{\hat \cv^2}\frac{1}{N(\tau)}
\left\{\sum_{i=1}^{N(\tau)-1}\left[(N(\tau)-i)^2\ln(\frac{\tau-T_{i}}{\tau-T_{i+1}})
     +i^2\ln(\frac{T_{i+1}}{T_i}) \right]\right. \nonumber \\
& & \hspace*{45mm}
 \left. +N(\tau)^2\left[\ln(\frac{\tau}{\tau-T_1})+\ln(\frac{\tau}{T_{N(\tau)}})-1 \right]
\right\}. 
\label{eq:AD}
\end{eqnarray}

\subsection{The Extended Lewis-Robinson Test for Non-Monotonic Trend}

Recall that the Lewis-Robinson type test for the time censored case
was based on the integral $\int_0^1 \tilde
V_{\tau,\hat\cv}^0(s)ds$. This test is suited for alternatives of
monotonic trend. Consider instead the expression \bq
 \label{vvint}
\int_0^{a} \tilde V_{\tau,\hat\cv}^0(s)ds - \int_{a}^1 \tilde
 V_{\tau,\hat\cv}^0(s)ds,
\eq 
where $0 \leq a \leq 1$. It is seen that $a=0$ in fact leads to the preferred test statistic (\ref{onestar}) for the Lewis-Robinson test (of course, $a=1$ gives the negative of the LR statistic (\ref{onestar})). 

A test based on (\ref{vvint}) will obviously have power to detect
non-monotonic trends where the trend in $[0,a \tau]$ and $[a
\tau,\tau]$ are in opposite directions.  Clearly, (\ref{vvint})
converges in distribution to $\int_0^{a} W^0(s)ds - \int_{a}^1
W^0(s)ds$, which is normally distributed with expectation 0 and
variance $1/12 - a^2(1-a)^2$ (see Appendix 2).  It follows from a
calculation in Appendix 2 that (\ref{vvint}), after a scaling to give
an asymptotically standard normal distribution under the null
hypothesis, can be written
\begin{equation}
\label{vvtest}
ELR = \frac{1}{\hat \cv} \cdot \frac{1}{\tau \sqrt{N(\tau)}\sqrt{(1/12)-a^2(1-a)^2}}
 \left\{ \sum_{i=1}^{N(\tau)} \left|T_i - a \tau\right| - \left(\frac{1}{2}-a(1-a)\right)\tau N(\tau)
\right\}.
\end{equation}

A disadvantage of the above test is that the value of $a$ has to be given. One possibility would of course be to allow an adaptive choice of $a$. This will, however, destroy the above distributional properties, and we will therefore not pursue this approach here. 

 \citeN{vaurio2} suggested on an ad hoc basis, and for the event
 censored case, a test statistic similar to (\ref{vvtest}) with $a=1/2$.

\subsection{Parameter Estimation}
\label{subsec:paramest}

If one  assumes the null hypothesis of HPP, then $\cv=1$
is known, and hence no estimation is needed in the use of
Corollary~\ref{cor1}.  If we more generally assume specific parametric
models for the event process, then the parameters $\mu,\sigma,\cv$ may be
estimated by maximum likelihood methods since they are functions of
the model parameters. In the case studies of Section~\ref{sec:case} we illustrate the
parametric estimation by fitting Weibull RPs to the
interevent times, taking into account also the censored time
at the end of the observation. Since the Weibull distribution is a
rather flexible distribution, the corresponding estimates of $\mu,\sigma$ and $\cv$ may be satisfactory also under the null hypothesis of RP when no parametric
assumptions are made. But strictly, when fitting Weibull distributions under
$H_0$, we test the null hypothesis that the events follow a Weibull
RP.

While this paper is basically about nonparametric trend testing, it should be
noted that fully parametric tests can be obtained by assuming a
parametric model for the original event process, where the null
hypothesis of RP refers to some parameter having a specific
value. A trend test can then be constructed by the likelihood ratio
method, see Section~\ref{aalenhusebye} for an example.  

When no distributional assumptions are made on the process,
obvious choices for estimators of $\mu$ and $\sigma$ are the sample mean $\hat \mu$ and sample standard deviation $\hat \sigma$
of the completely observed interevent times. These estimators are
consistent as $\tau \rightarrow \infty$ (see Appendix~1), but have the
disadvantage of not utilizing the censored times $\Y$ at the end of the
observation period. The corresponding estimator of $\cv$ is $\hat \cv = \hat \sigma/\hat \mu$. 

Alternative estimators which involve the censored time $\Y$ may be derived from standard renewal process theory. Again we refer to Appendix~1 for justification of the following estimators,
\begin{equation}
\tilde \mu =\frac{\tau}{N(\tau)}, \; \; \;
\tilde \sigma^2 =      \frac{1}{N(\tau)} \left\{ \sum_{i=1}^{N(\tau)} X_i^2 + (\tau - T_{N(\tau)})^2 \right\} - \tilde \mu^2, \; \; \; \tilde \cv = \tilde \sigma/\tilde \mu.
\end{equation}

Another variance estimator (see  Appendix~1 for its verification) is
\begin{equation}
\sigma^{*2}=\frac{1}{2(N(\tau)-1)}\sum_{i=1}^{N(\tau)-1}(X_{i+1}-X_i)^2.
\label{varest}
\end{equation}
The potential advantage of this estimator is
that it tends to be smaller than  $\hat \sigma^2$ and $\tilde
\sigma^2$ under alternatives with positive dependence between
subsequent interevent times. This makes the estimated $\cv$ become
smaller, which leads to larger (absolute) values of the test
statistics and hence higher rejection probability under alternatives
of monotonic trend, see for example \citeN{vaurio2}. We will, however,
in our simulation and data examples use $\hat \sigma$ or $\tilde
\sigma$ and not $\sigma^*$, due to apparent less satisfactory
significance level properties, as experienced in simulations.

\section{Tests for Trend in Multiple Processes}
\label{sec:several}

Suppose now that $m >1$ similar processes are observed. Under the assumption that the processes are stochastically independent it may be of interest to test the null hypothesis that they all have no trend. One possible formulation of the null hypothesis is to let $H_0$ state that all the $m$ processes are independent RPs, but that they are not necessarily identically distributed. A stronger null hypothesis would be to state that the $m$ processes are independent RPs with the same distribution of the interevent times. We will below mostly stick to the former interpretation, but will consider the latter hypothesis in the example of Section~\ref{aalenhusebye}.

Construction of the tests is based on the following fact, which we state as a lemma:

\begin{lemma}
\label{lemma1}
 Let $W_1^0, W_2^0,\ldots,W_m^0$ be independent Brownian bridges and let $a_1,a_2,\ldots,a_m$ be real numbers with $\sum_{j=1}^m a_j^2 = 1$.  Then 
  \[      
      W^0 =  \sum_{j=1}^m a_j W_j^0
      \]
      is a Brownian bridge.
      \end{lemma}
      
      \noindent \textit{Proof:} By linearity it is clear that $W^0$ is a Gaussian process with expectation $0$. The result follows by a straightforward calculation of the covariance function.  
      
      \bigskip

 Let $\tau_j$, $\mu_j$, $\sigma_j$ and $\cv_j$ be,
respectively, the censoring time, mean, standard
deviation and coefficient of variation corresponding to process $j$, $j=1,\ldots,m$.  Let further $\A_1,\ldots,\A_m$ be random variables where $\A_j$ depends on the data from process $j$ only, and assume that $\A_j \stackrel{p}{\rightarrow} a_j$, $j=1,\ldots,m$, where the $a_j$ are constants with  $\sum_{j=1}^m a_j^2 = 1$. Then from Lemma~\ref{lemma1}, Corollary~\ref{cor1} and  the already cited 'converging together lemma' it follows that
 \begin{equation}
\sum_{j=1}^m A_j \tilde V_{\tau_j,\cv_j}^0(s)
=\sum_{j=1}^m A_j \frac{1}{\cv_j} \frac{N_j(s\tau_j)-sN_j(\tau_j)}{\sqrt{N_j(\tau_j)}}  \;   \Rightarrow \;  W^0 \; \mbox{ as } \tau_j \rightarrow \infty, \; j=1,\ldots,m.
\label{eq:mbbtime}
\end{equation}

Depending on the choice of weights $A_j$, this can lead to different
generalizations of the tests in Section~\ref{sec:classof}. One way of constructing tests will be to perform the same transformations as in Section~\ref{sec:classof} to the left hand side of (\ref{eq:mbbtime}). This is a straightforward operation for the Lewis-Robinson type tests, but for the other types of tests, the derivation of the test statistics will be more cumbersome. For these  we might therefore instead consider linear combinations of the tests for single processes. 

\vekk{ 
Notice that under the null hypothesis $H_0$ of independent RPs, the $m$
processes are allowed to have different parameters, so that separate estimates $\hat\mu_j$,
$\hat\sigma_j$ and $\hat \cv_j$ are calculated for each process $j=1,\ldots,m$. If, on the other hand, it is reasonable to assume the same distribution in all processes, then common estimators of $\mu$, $\sigma$ and $\cv$ based on the data from all processes can be calculated. This may also be the preferred action for relatively small data sets, see the example in subsection~\ref{aalenhusebye}. 

We will consider in some detail the Lewis-Robinson type test, since the derivation leads to a natural extension of the Lewis-Robinson test for the event censored case which goes back to \citeN{lewisrobinson}. 
} 

\subsection{Lewis-Robinson Type Test for $m$ Processes}

By the same arguments as in Section~\ref{subsec:lr}, and with the assumption on the weights given above, the following
statistic will be asymptotically standard normally distributed under $H_0$,
\begin{equation}
LR ^m= -\sqrt{12} \int_0^1 \sum_{j=1}^m A_j \tilde V_{\tau_j,\cv_j}^0(s)ds
 =\sum_{j=1}^m A_j\frac{1}{\hat \cv_j} \cdot \frac{\sqrt{12}}{\tau_j\sqrt{N_j(\tau_j)}}
 \left[\sum_{i=1}^{N_j(\tau_j)} T_{ij} - \frac{N_j(\tau_j)}{2}\tau_j
\right].
\label{eq:mLR}
\end{equation}
Here $T_{ij}$ denotes the time until failure number $i$ in process $j$,  $i=1,\ldots,N_j(\tau_j)$, $j=1,\ldots,m$. 

Different choices of the weights $A_j$ will lead to different tests.  For instance, $A_j=1/\sqrt{m}$, $j=1,\ldots,m$ will mean equal weighting of the information from each
process. This might, however, not be an optimal choice in cases where the processes have been observed for different lengths of time, or if there is a large variation in the number of events per process.

For the Poisson process case, \citeN{kvaloeylindqvist} suggested to generalize the  Laplace test for a single process to a test statistic based on standardizing the sum  $\sum_{j=1}^m \sum_{i=1}^{N_j} T_{ij}$. In the more general situation considered here, the form of the coefficients on the right hand side of (\ref{eq:mLR}) suggests the use of weights $A_j$  such that
\begin{equation}
\label{LRA}
   A_j \; \propto \;
\hat \cv_j  \tau_j\sqrt{N_j(\tau_j)}.
\end{equation}
Suppose now that the $\tau_j$ tend to infinity in such a manner that, for a $\tau$ tending to infinity, $\tau_j/\tau \rightarrow c_j$ for positive constants $c_j$, $j=1,\ldots,m$. Since the $N_j(\tau_j)/\tau_j \rightarrow 1/\mu_j$ a.s. and $\hat \cv_j \stackrel{p}{\rightarrow} \cv_j$, we have
\begin{equation}
\label{twostar}
A_j=\frac{\hat \cv_j \tau_j \sqrt{N_j(\tau_j)}}{\sqrt{\sum_{k=1}^{m}\hat \cv_k^2 \tau_k^2N_k(\tau_k)}}
= \frac{\hat \cv_j (\tau_j/\tau)^{3/2} \sqrt{N_j(\tau_j)/\tau_j}}{\sqrt{\sum_{k=1}^{m}  \cv_k^2 (\tau_k/\tau)^{3} N_k(\tau_k)/\tau_k}}
\stackrel{p}{\rightarrow}
\frac{\cv_j c_j^{3/2}/ \sqrt{\mu_j}}{\sqrt{\sum_{k=1}^{m}\cv_k^2 c_k^3/\mu_k}} \equiv a_j.
\end{equation}
Clearly, $\sum_{j=1}^m a_j^2=1$, so the  statistic (\ref{eq:mLR}) will converge to a standard normal
distribution under the null hypothesis $H_0$. 

Inserting the weights $A_j$ from (\ref{twostar}) and rearranging we can write the test statistic (\ref{eq:mLR}) as
\begin{equation}
\label{lrm}
LR ^m= \frac{\sqrt{12}}{\sqrt{\sum_{k=1}^{m}\hat \cv_k^2 \tau_k^2N_k(\tau_k)}}
 \sum_{j=1}^m  
 \left[\sum_{i=1}^{N_j(\tau_j)} T_{ij} - \frac{N_j(\tau_j)}{2}\tau_j
\right].
\end{equation}

\citeN{Lawless2012tmt} considered a
similar test statistic for the time censored case,  but under the slightly different
null hypothesis that all the $m$  processes have constant rate
functions, and with asymptotics as $m \rightarrow \infty$.  Let $U_j=\sum_{i=1}^{N_j(\tau_j)}T_{ij}-N_j(\tau_j)\tau_j/2$ for $j=1,\ldots,m$.  
The test statistic of what they named \textit{the generalized Laplace test} is
\[
GL=\frac{\sum_{j=1}^mU_j}{\sqrt{\sum_{j=1}^mU_j^2}},
\]
which under the null hypothesis is asymptotically standard normal as
$m\rightarrow\infty$.

\subsection{Other Tests for $m$ Processes}
\label{subsec:m}

For the other tests considered in Section~\ref{sec:classof} it is in principle possible to replace the $\tilde
V_{\tau,\cv}^0(s)$ by $\sum_{j=1}^m A_j \tilde V_{\tau_j,\cv_j}^0(s)$ and
apply the same operations as for the case $m=1$. This corresponds to what we did for the Lewis-Robinson test in the previous subsection, but here things are easy  due to the linearity of integrals. This also applies to the extended Lewis-Robinson test. 
For the remaining tests it is not straightforward, however, to derive explicit
expressions for the test statistics, and it is neither clear what would be
the best weights to use.  The problem associated with the Cram\'{e}r-von Mises and Anderson-Darling
tests are of course that the integrand is a square, while for the
Kolmogorov-Smirnov test the various processes are mixed together
before taking the absolute value, making tractable expressions
impossible.

Another possibility for these last mentioned tests would therefore be to use
(weighted) sums of the individual test statistics to define the new
test statistics. Such an approach requires, on the other hand, the distributions of sums or linear combinations of the limiting
distributions for the single process cases. These may be determined by
simulations or, for larger $m$, by normal approximations. Note also
that \citeN{scholzstephens} have considered the distribution of sums
of independent Anderson-Darling statistics. 

For such linear combinations there are no obvious choices for the weights given to each process. A
reasonable choice under the assumption of the same interevent
distribution in all processes would be to let the weights be
proportional to $\tau_j$.  Otherwise, it may be tempting to use
weights like (\ref{LRA}), hence taking into account the length of
observation of each process as well as the number of observed events
and the coefficient of variation of the interevent times.  A problem would then of course be that these weights are random, making exact simulation of the
distribution under the null hypothesis impossible.  

In practice we have found that the normal approximation works fairly well
for the Cram\'{e}r-von Mises test, but less well for the Anderson-Darling
test due to the very skew distribution of the Anderson-Darling statistic.

\section{Simulation Study}
\label{sec:sim}

We have done various simulations to study and compare the properties
of the tests. When we report results for single processes we do not include the
Cram$\acute{\mbox{e}}$r-von Mises test as this test had less power
than the Anderson-Darling test, while for several
processes we do not include the Anderson-Darling test as the
Cram$\acute{\mbox{e}}$r-von Mises test had better level properties in
this case as discussed in Section~\ref{subsec:m}. For the
extended Lewis-Robinson test we chose $a=1/2$ in (\ref{vvtest})
and we only report this test for non-monotonic trend as
it has inferior power against monotonic trends.

In the reported simulations we estimated rejection probabilities by
simulating 100 000 data sets for each choice of model and parameter
values, and recorded the relative number of rejections of each
test. The standard errors of the simulated rejection probabilities are then $\leq
0.0016$. All simulations were done in R. The nominal significance
level was set to 5\%.

To simulate data with trend, we used the trend-renewal
processes (TRP) \cite{leh} which in short is defined as follows:  
Let $\lambda(t)$ be a non-negative function
defined for $t\geq 0$ and let $\Lambda(t)=\int_0^t\lambda(u)du$. Then
the process $T_1,T_2,\ldots$ is a TRP with \textit{trend function} $\lambda(t)$ and \textit{renewal distribution} $F$, if
$\Lambda(T_1),\Lambda(T_2),\ldots$ is an RP with interevent times having the distribution~$F$.

The RP, the non-homogeneous Poisson process (NHPP) and the HPP are all
special cases of the TRP. For example, if the trend function is
constant, then the TRP is an RP, while if the distribution $F$ is the
unit exponential distribution, then the TRP is an NHPP with intensity
function $\lambda(t)$.  The trend in a TRP is hence governed by the
trend function $\lambda(t)$, and by letting the distribution $F$ be
any positive-valued distribution, we are left with a large class of
processes with trend.  In our simulations we will use
parameterizations of the TRP where the renewal distribution $F$ is a Weibull-distribution and
the trend function is either of so called power law  or bath tube type, see
Section~\ref{subsec:simonepower} below.

\subsection{One Process - Level Properties}
\label{subsec:simonelevel}

\begin{figure}[!hbt]
\begin{center}
\includegraphics[width=0.44\linewidth]{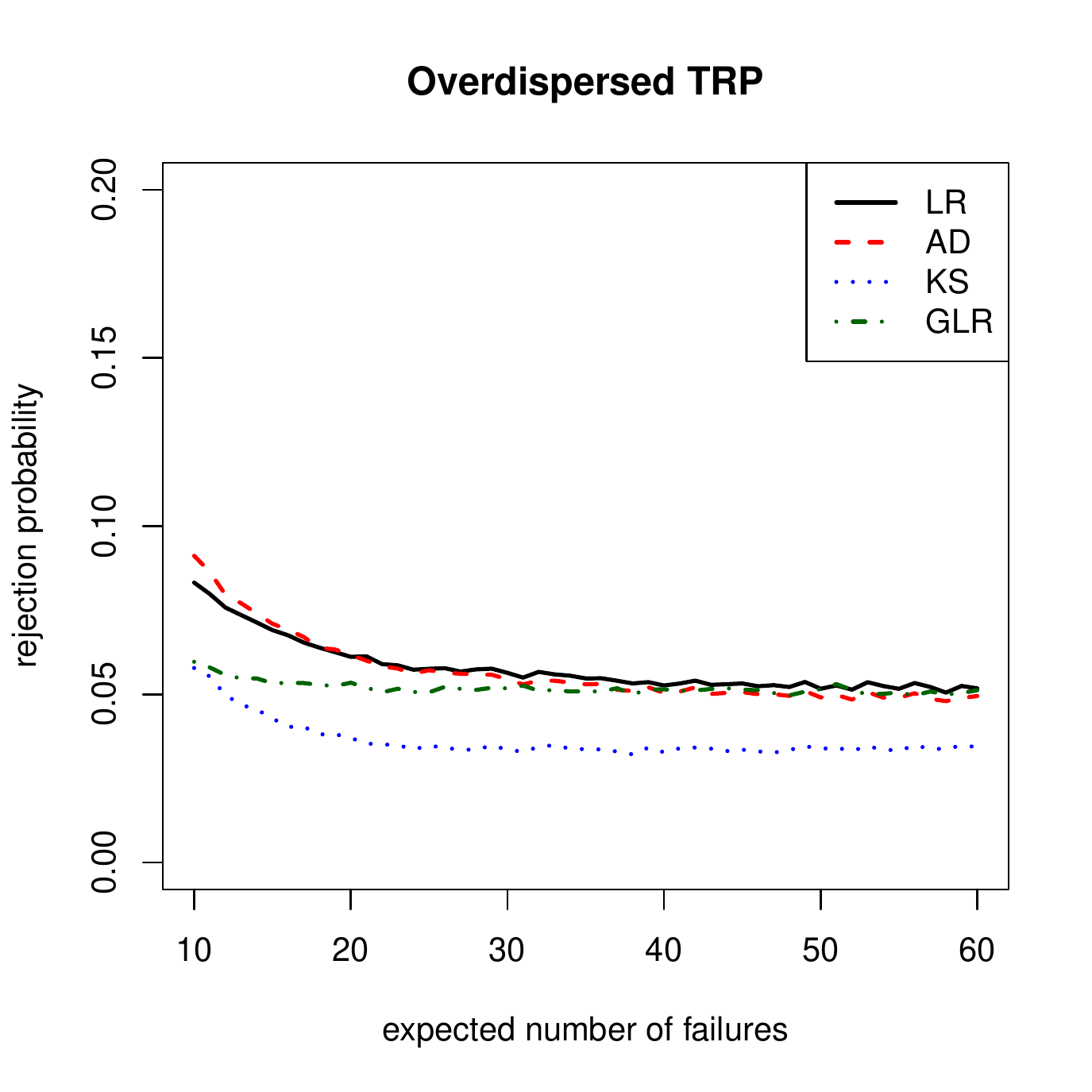}
\includegraphics[width=0.44\linewidth]{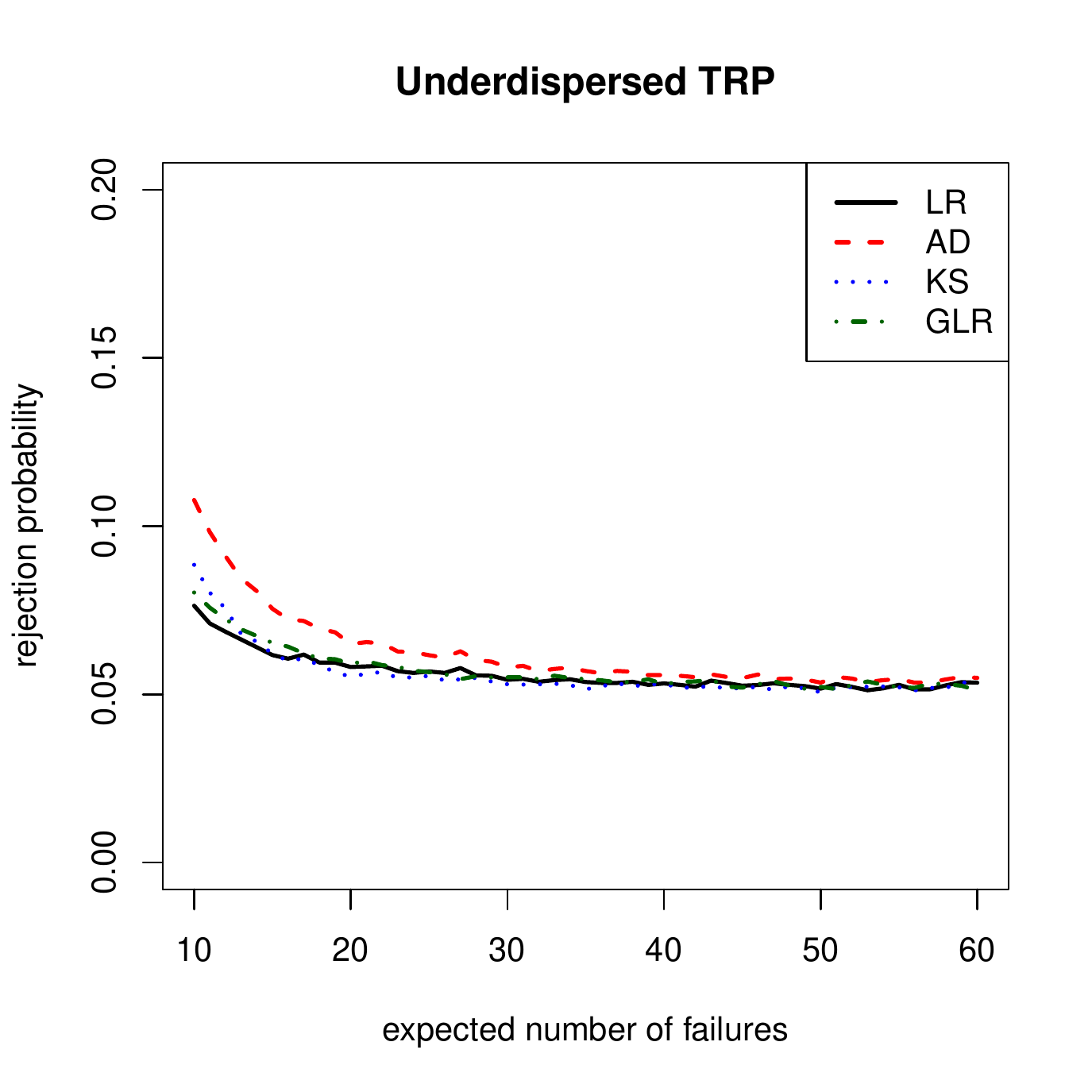}
\end{center}
\caption{Simulated level properties as a function of expected number
  of events, with data generated from Weibull RPs with shape
parameters respectively 0.75 (overdispersed RP) and 1.5
(underdispersed RP). Abbreviations: LR = Lewis-Robinson,
KS = Kolmogorov-Smirnov, AD = Anderson-Darling, ELR = Extended
Lewis-Robinson test.}
\label{fig:nivaafig}
\end{figure}

First the level properties of the tests were studied by generating
data sets from Weibull RPs with shape parameters respectively 0.75 and
1.5, corresponding respectively to a process which is overdispersed
and a process which is underdispersed relative to an HPP.  In
Figure~\ref{fig:nivaafig} the simulated level of the tests for 
systems with the expected number of events ranging from 10 to 60 are reported.

The tests mostly have adequate level properties, but  all tests
are a bit non-conservative for small samples in the underdispersed
case, while the  Kolmogorov-Smirnov test is too conservative in the
overdispersed case. \\

\subsection{One Process - Power Properties}
\label{subsec:simonepower}

Data sets with a monotonic trend were generated by simulating data
from TRPs with the renewal distribution being Weibull and
the trend function $\lambda(t)$ being of the power law form $\lambda(t)=b t^{b-1}$. 
The rejection probability as a function of $b$ was simulated, where $b<1$
corresponds to a decreasing trend, $b=1$ corresponds to no trend
and $b>1$ corresponds to an increasing trend.
\begin{figure}[!hbt]
\begin{center}
\includegraphics[width=0.44\linewidth]{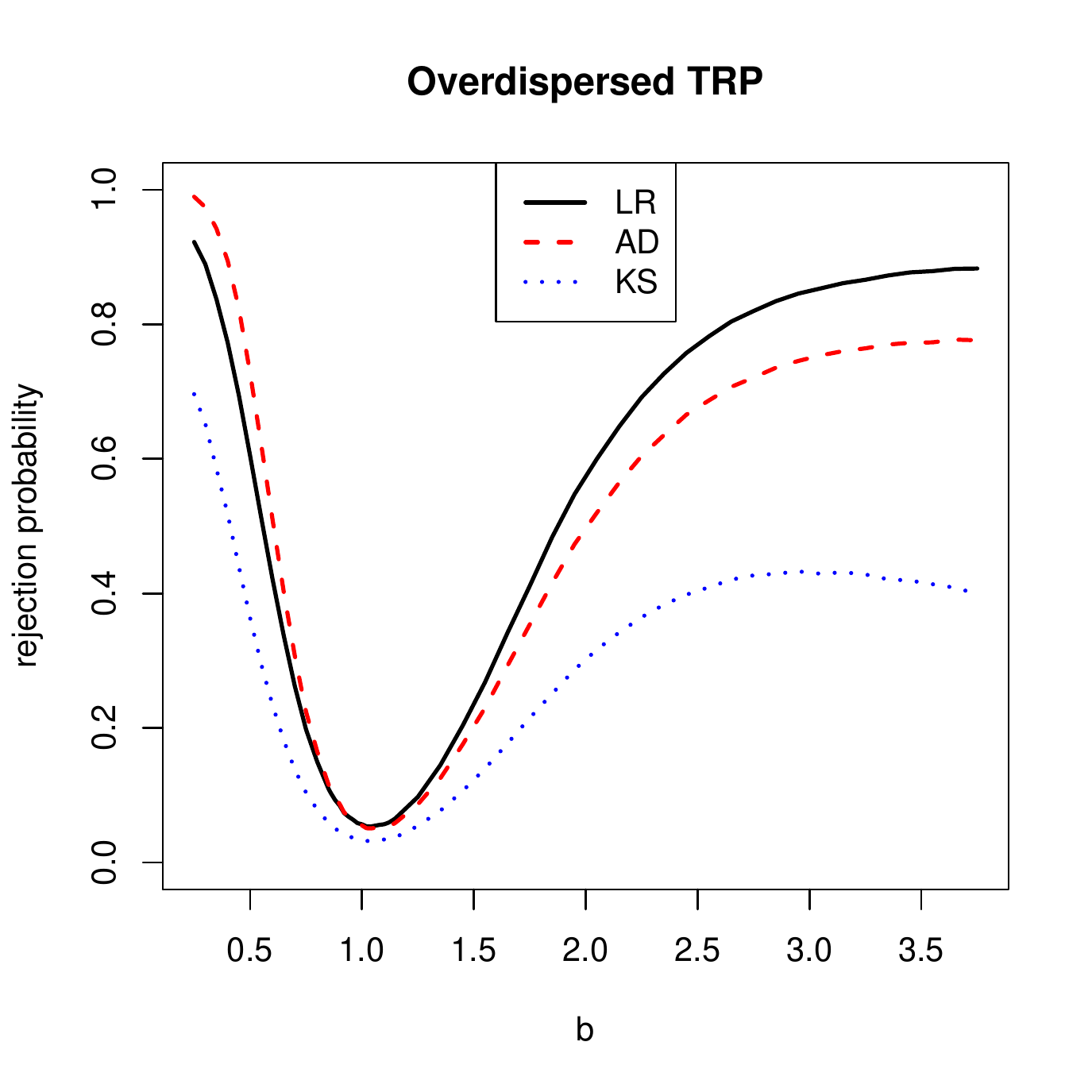}
\includegraphics[width=0.44\linewidth]{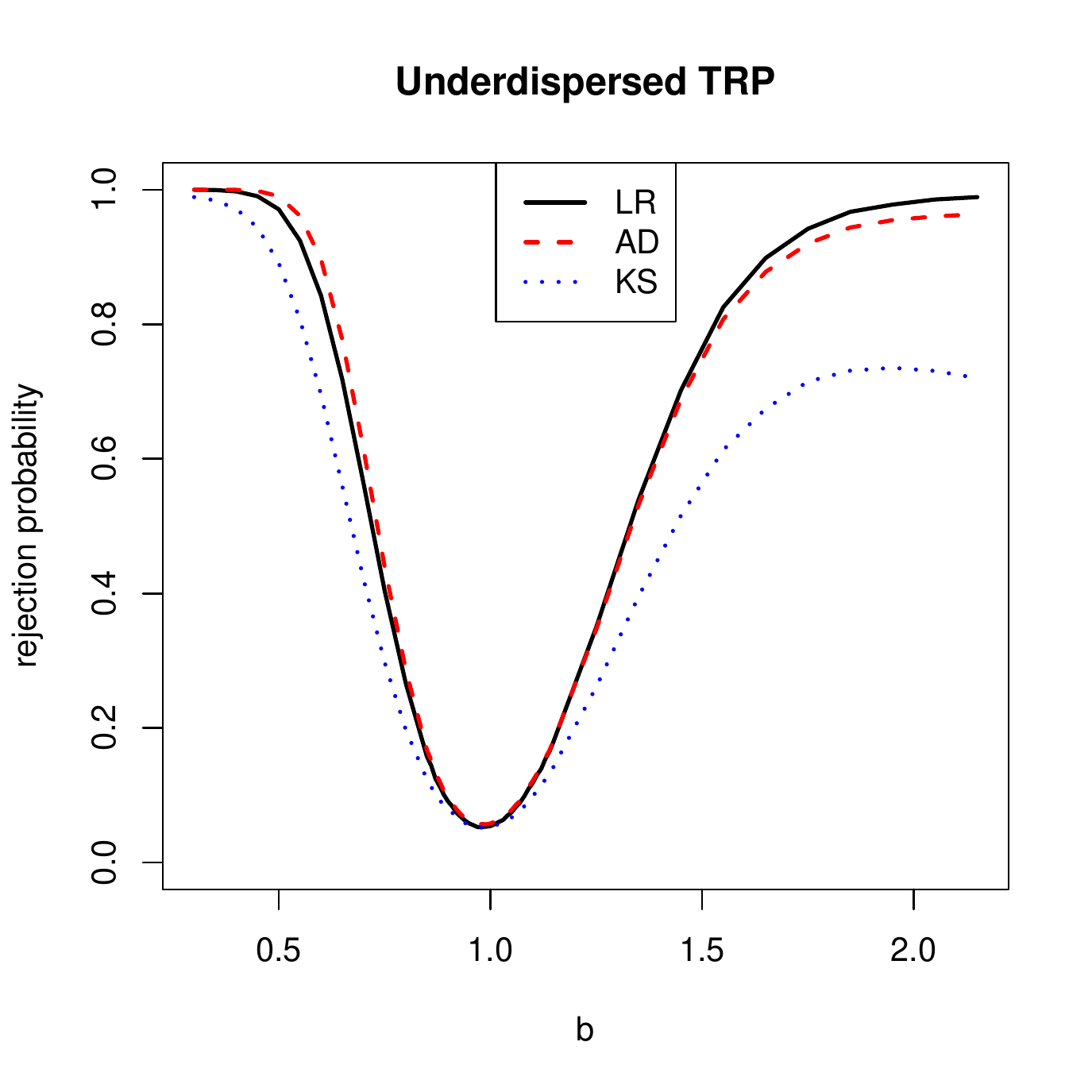}
\end{center}
\caption{Simulated power properties as a function of trend parameter
  $b$, with data simulated from TRPs with trend function $b t^{b-1}$
  and Weibull renewal distribution with shape 
parameters respectively $\beta=0.75$ (overdispersed TRP) and $\beta=1.5$
(underdispersed TRP). The censoring time is adjusted such that the expected number of
 failures is 30. Abbreviations: LR = Lewis-Robinson, KS =
 Kolmogorov-Smirnov, AD = Anderson-Darling.}
\label{fig:powerfig}
\end{figure}
Two different values of the shape parameter
$\beta$ of the Weibull renewal distribution were considered, $\beta=0.75$ and
$\beta=1.5$. The censoring times were adjusted such that the expected number of
 failures was 30. The results are displayed in Figure~\ref{fig:powerfig}.
We see in this figure that the Anderson-Darling test is the most powerful
test against decreasing trend, but is a
bit less powerful than the Lewis-Robinson test for increasing
trend. The  Kolmogorov-Smirnov test is less powerful than the other tests. \\

\begin{figure}[!htb]
\begin{center}
\includegraphics[keepaspectratio=true,width=3.6in]{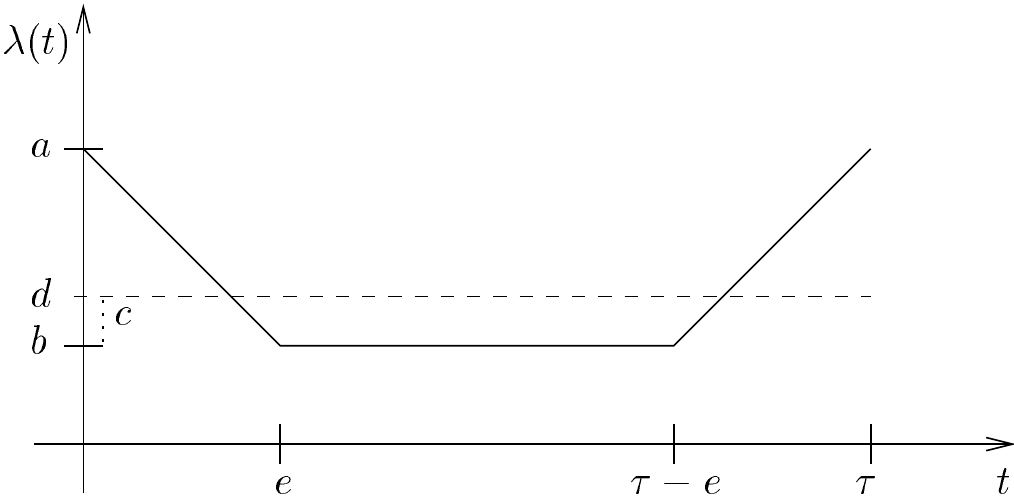}
\end{center}
\caption{Bathtub-shaped trend function.}
\label{fig:badefig}
\end{figure}

Data sets with a bathtub trend were generated by simulating data
from TRPs with trend function $\lambda(t)$ on the form displayed
in Figure~\ref{fig:badefig}.
Here $d$ is the average of  $\lambda(t)$
over $[0,\tau]$. The degree of bathtub shape can be
expressed by the parameter $c$, with $c=0$ corresponding to a
horizontal line (no trend).

\begin{figure}[!hbt]
\begin{center}
\includegraphics[width=0.44\linewidth]{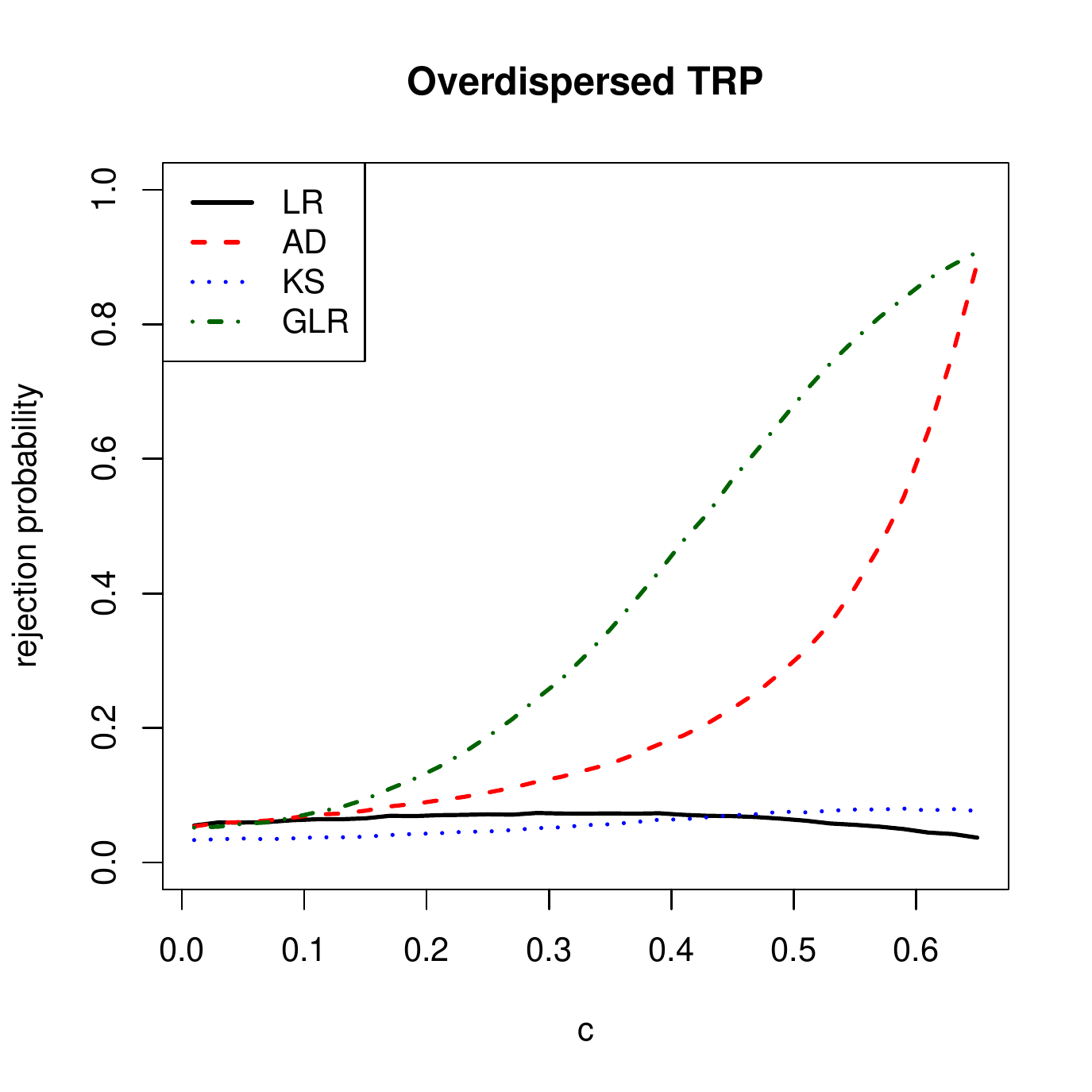}
\includegraphics[width=0.44\linewidth]{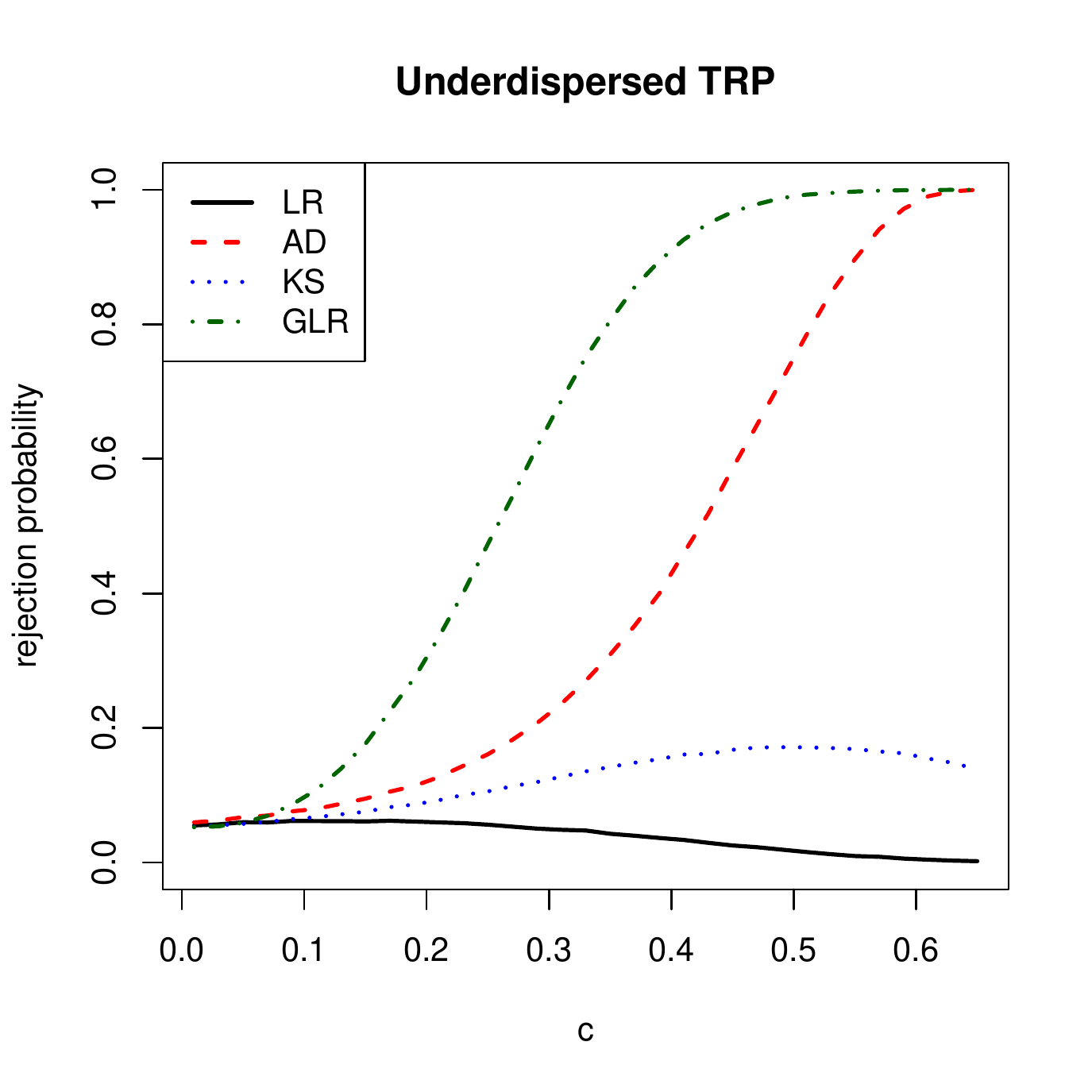}
\end{center}
\caption{Simulated power properties as a function of trend parameter
  $c$, with data simulated from  TRPs with bathtub trend function (Figure~\ref{fig:badefig}),  Weibull renewal distribution with shape
parameter $\beta=0.75$ (overdispersed) and $\beta=1.5$ (underdispersed TRP),  and expected number of failures in each phase
equal to 20. Abbreviations: LR = Lewis-Robinson, KS =
Kolmogorov-Smirnov, 
AD = Anderson-Darling, ELR = Extended Lewis-Robinson test}
\label{fig:badetrendfig}
\end{figure}

The rejection probability as a function of $c$ was simulated with
$e$ and $\tau$ in each case set to values such that the 
expected number of failures in each phase (decreasing, no,
increasing trend) were equal to 20. The shape parameter of the Weibull renewal  distribution was set to respectively $\beta=0.75$ and $\beta=1.5$. The
results are displayed in Figure~\ref{fig:badetrendfig}.

We see in Figure~\ref{fig:badetrendfig} that the extended
Lewis-Robinson test and the Anderson-Darling test have the ability to
detect 
this non-monotonic trend, while the other tests have no power in
such cases.  Not surprisingly, the trend is easier to detect in
the underdispersed case. The extended
Lewis-Robinson test with $a=1/2$ (\ref{vvtest})  is
by its construction particularly well suited for picking up
non-monotonic trends which are symmetric around the mid-point of the
observation interval, $\tau/2$, as we have in this case.   \\

\subsection{Several Processes }

When considering several processes, the number of processes is one of
the important factors for the behavior of the tests. We show here
some simulations which illustrate power and level properties for the
test with different number of processes. In this setting with several processes
the generalized Laplace test also applies. 

\begin{figure}[!hbt]
\begin{center}
\includegraphics[width=0.44\linewidth]{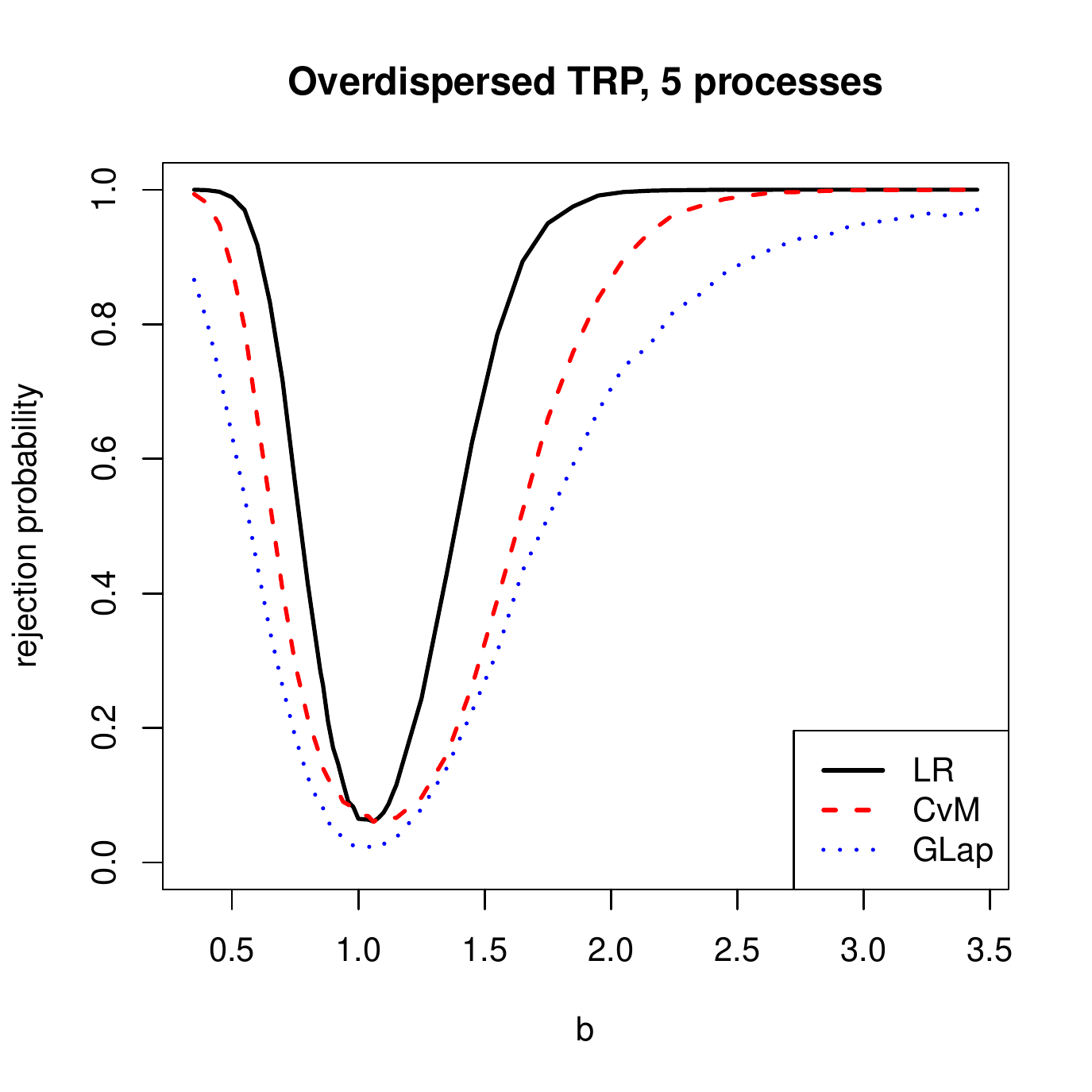}
\includegraphics[width=0.44\linewidth]{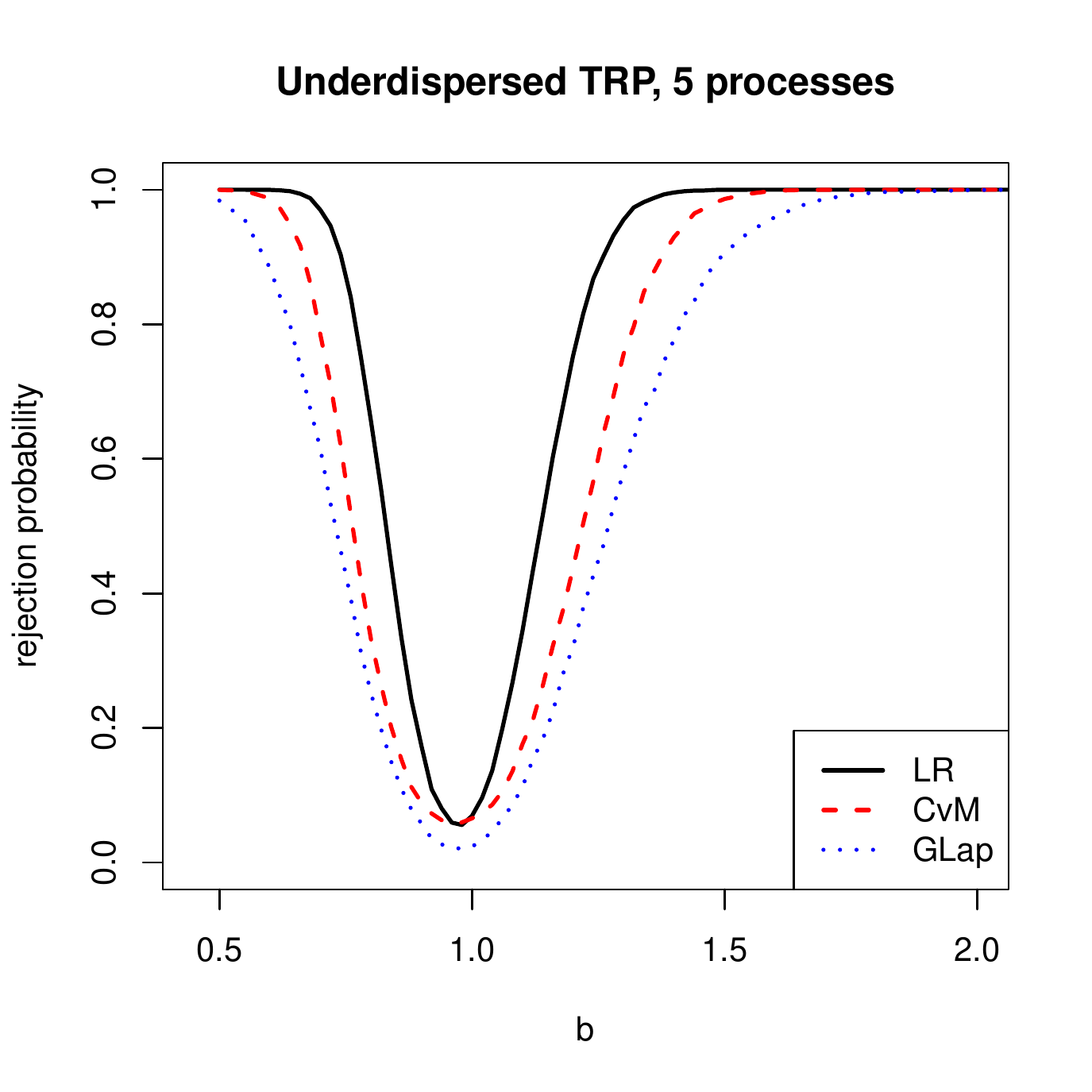}
\end{center}
\caption{Simulated power properties as a function of trend parameter
  $b$, with data simulated from 5 TRPs with trend function $b
  t^{b-1}$ and Weibull renewal distribution with shape
parameters respectively $\beta=0.75$ (overdispersed TRP) and $\beta=1.5$
(underdispersed TRP). The censoring time is adjusted such that the expected number of
 failures in each process is 20. Abbreviations: LR = Lewis-Robinson,
 CvM = Cram$\acute{\mbox{e}}$r-von Mises,  GL = Generalized Laplace Test.}
\label{fig:mpowerfig1}
\end{figure}

\begin{figure}[!hbt]
\begin{center}
\includegraphics[width=0.44\linewidth]{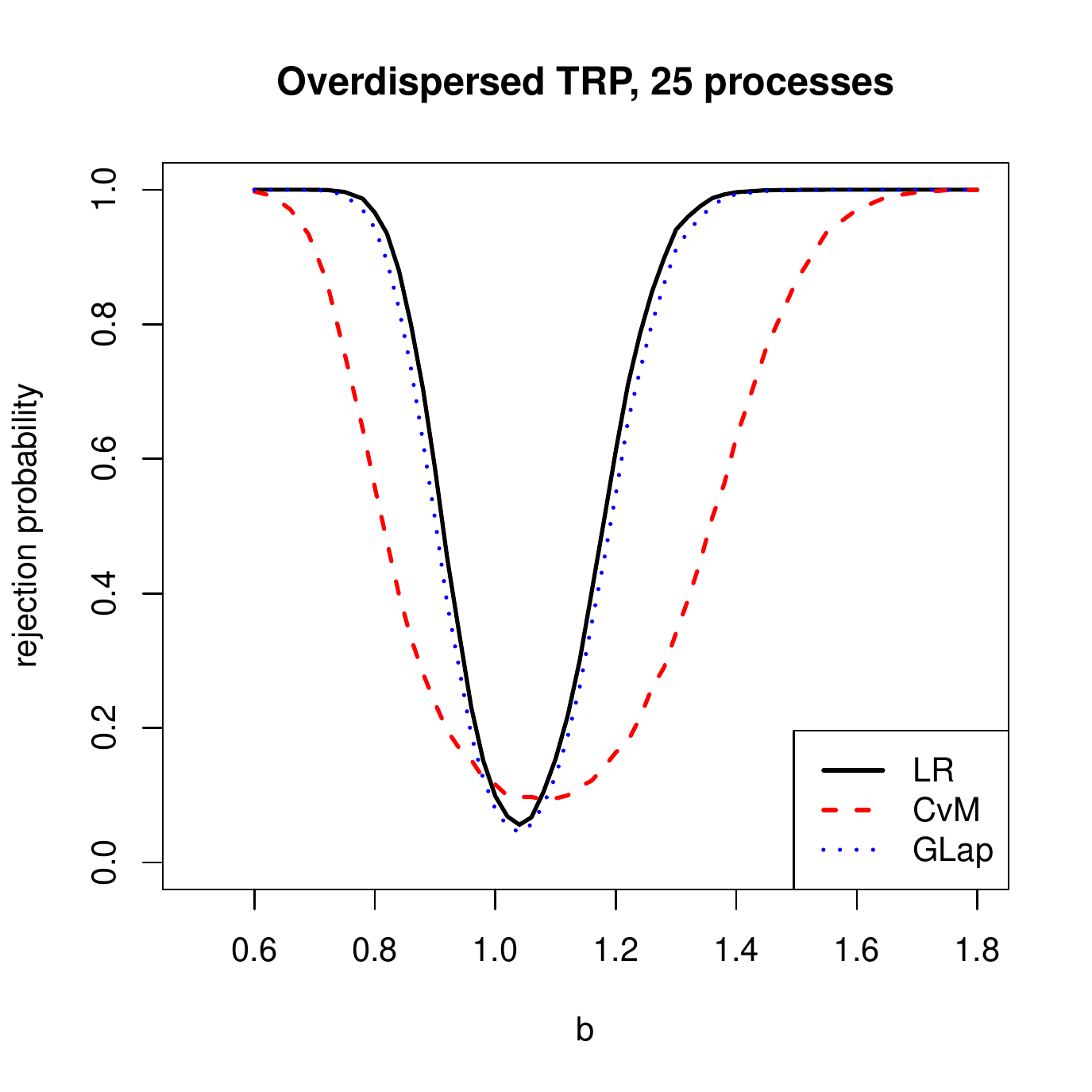}
\includegraphics[width=0.44\linewidth]{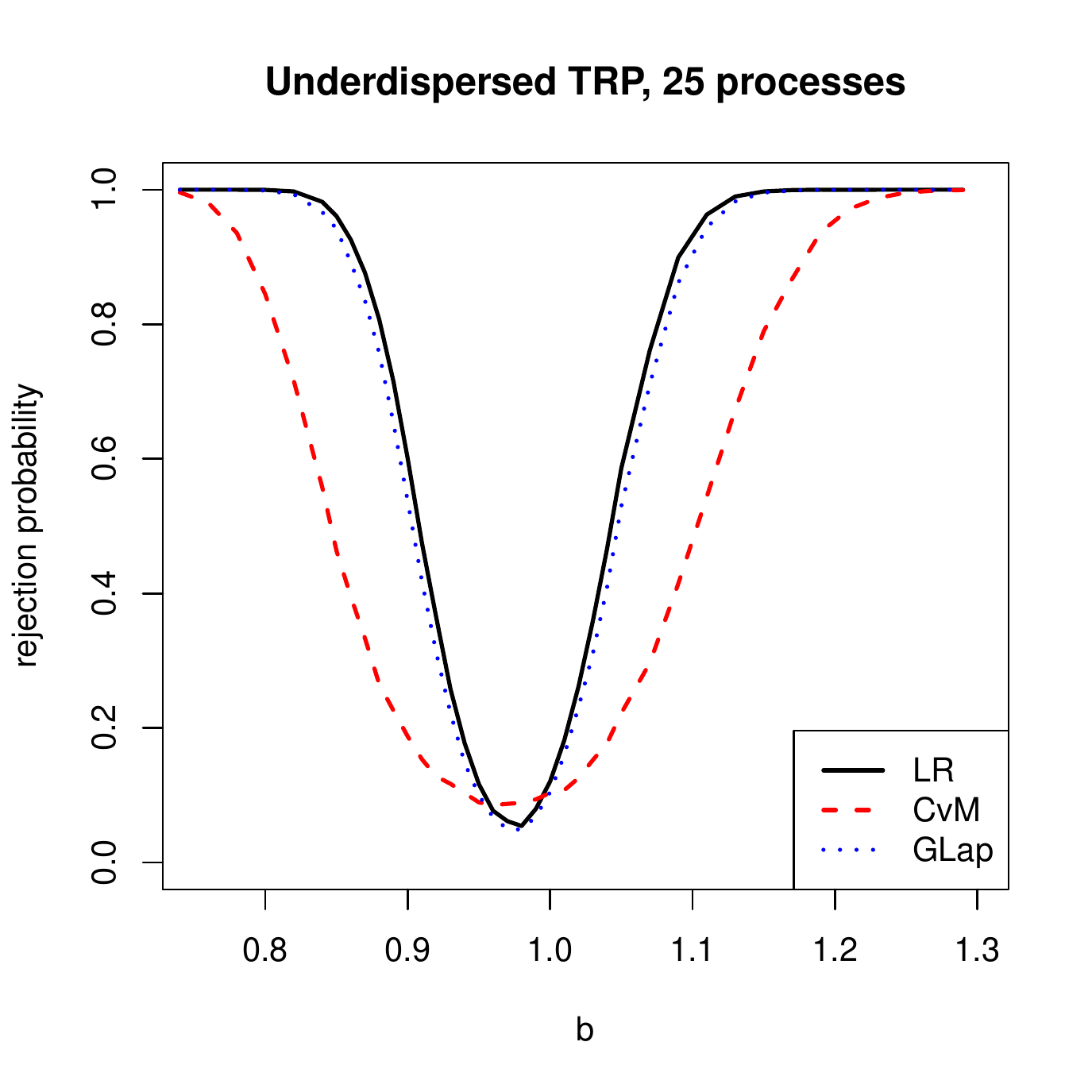}
\end{center}
\caption{Simulated power properties as a function of trend parameter
  $b$, with data simulated from 25 TRPs with trend function $b
  t^{b-1}$ and  Weibull renewal distribution with shape
parameters respectively $\beta=0.75$ (overdispersed TRP) and $\beta=1.5$
(underdispersed TRP). The censoring time is adjusted such that the expected number of
 failures in each process is 20. Abbreviations: LR = Lewis-Robinson,
 CvM = Cram$\acute{\mbox{e}}$r-von Mises,  GL = Generalized Laplace Test.}
\label{fig:mpowerfig2}
\end{figure}

Figures~\ref{fig:mpowerfig1} and \ref{fig:mpowerfig2} show power
properties  for cases with respectively 5 and 25 processes
and with censoring time chosen such that the expected number of
events in each process is 20. Simulations with other expected number
of failures showed similar behavior, just with lower or higher power
depending on whether the expected number of failures was lower or
higher. These simulations show that the Lewis-Robinson type test has
the best power properties in these monotonic trend cases. We also
notice that the generalized Laplace test is very similar to the
Lewis-Robinson test in the case with 25 processes. 


\clearpage
\section{Case Studies}
\label{sec:case}

\subsection{Load-Haul-Dump Machine Data (Kumar et al., 1989)}
\label{LHDdata}

\begin{figure}[!hbt]
\begin{center}
\includegraphics[width=0.45\linewidth]{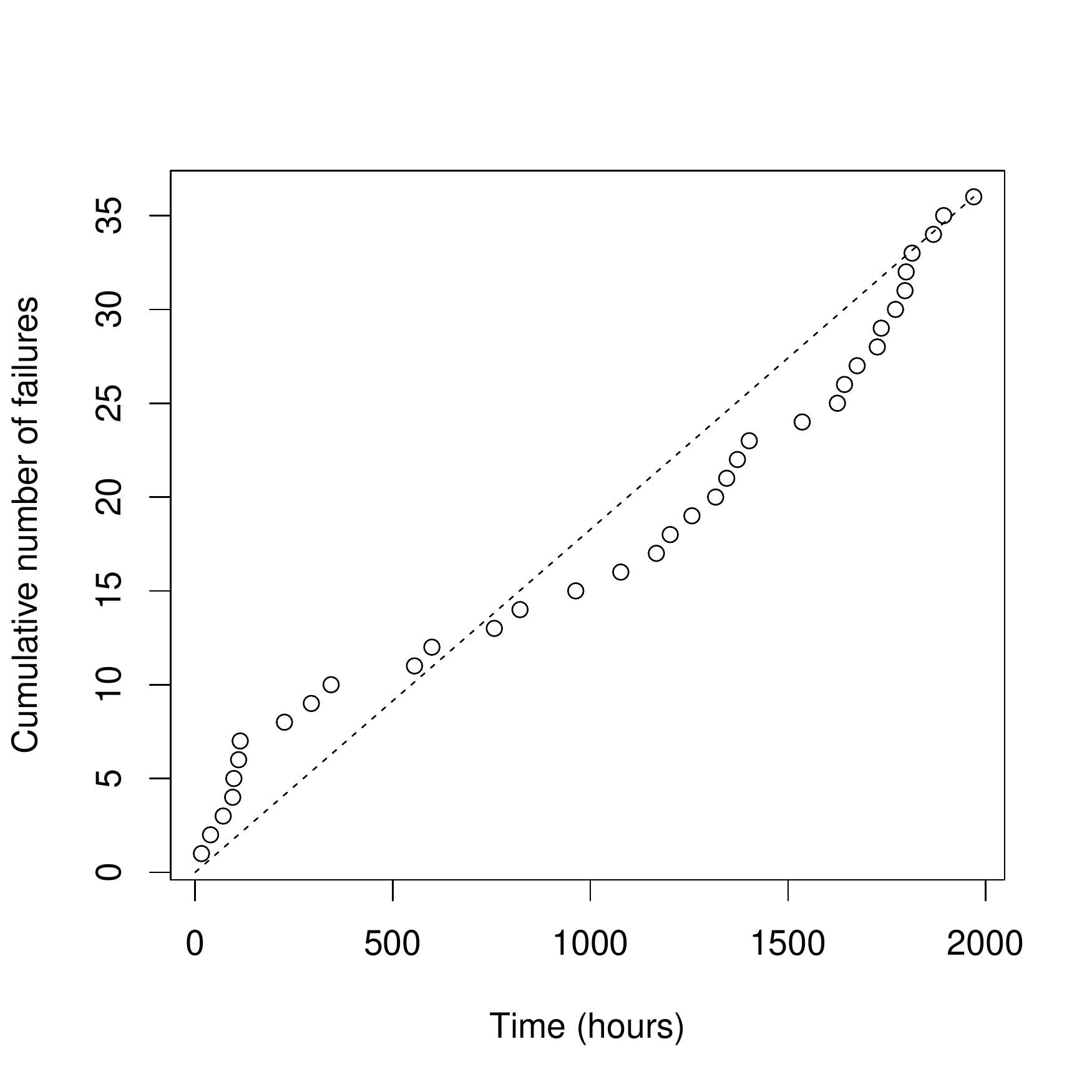}
\includegraphics[width=0.45\linewidth]{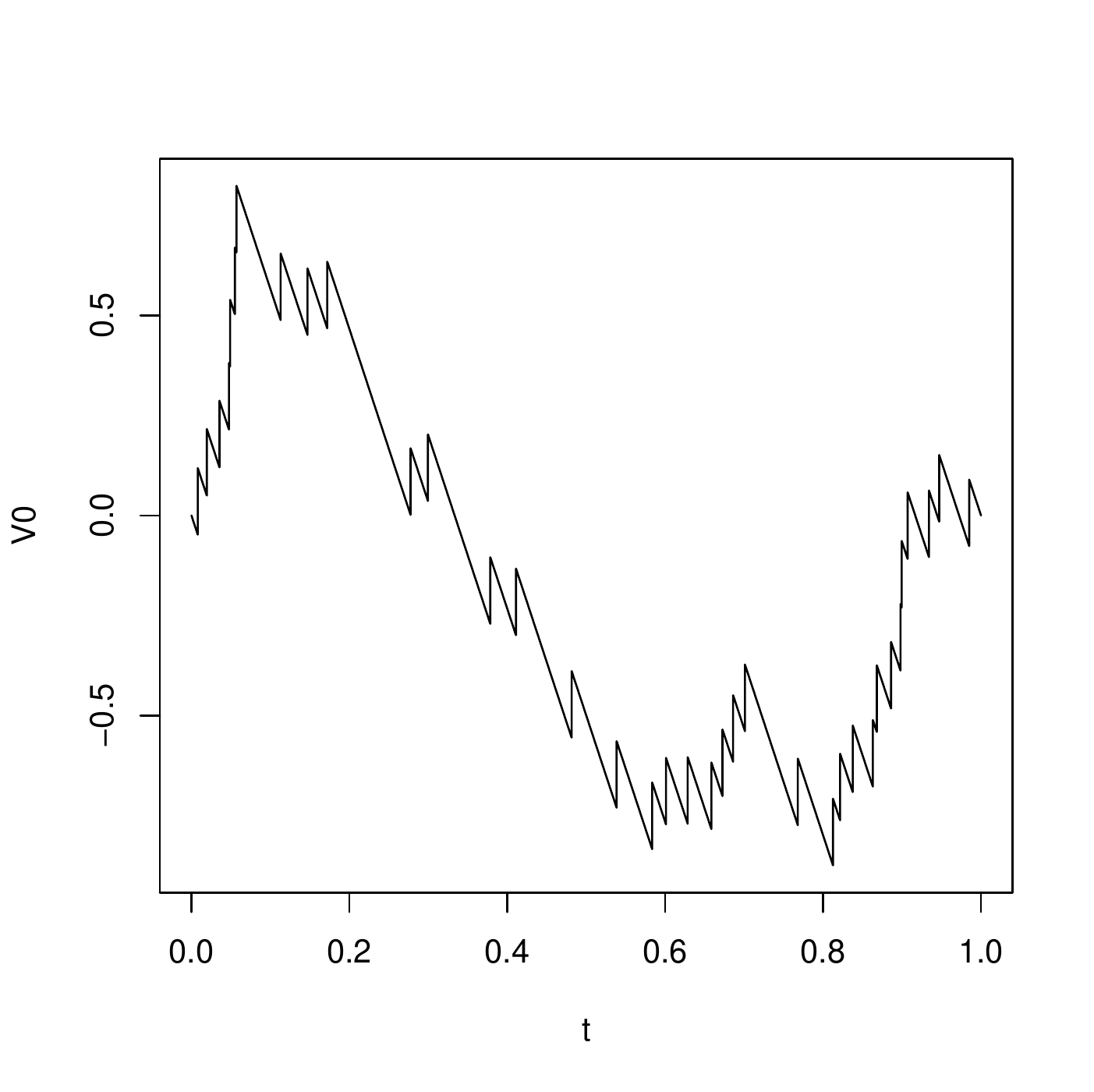}
\end{center}
\caption{Load-haul-dump data. Left: Plot of $N(t)$ for $0 \le t \le 2000$. Right: Plot of the process $\tilde V^0_{\tau,1}(s)$ for $0 \le s \le 1$.}
\label{fig:LHDAdata}
\end{figure}

\citeN{kkg} reported failure data for a load-haul-dump machine operating
in a Swedish mine. For the purpose of this example we considered the data to be time censored at $\tau=2000$ hours. The recorded failure times of the machine up
to this time are reported in Table~\ref{tab:LHDdata}, and a plot
of the observed process $N(t)$ for $0 \le t \le 2000$ is given in the left panel of
Figure~\ref{fig:LHDAdata}. The plot seems to indicate a
non-monotonic trend, apparently in the form of a
bathtub trend. 

For illustration we also show, in the right panel of
Figure~\ref{fig:LHDAdata}, a plot of the function $\tilde V^0_{\tau,1}(s)$
for $0 \le s \le 1$. This is the transformed and tied down version of
$N(t)$, and should, if the null hypothesis holds, be close to a
Brownian bridge. However, this plot too indicates a non-monotonic trend with
an upward deviation in the first part and a downward deviation in the
second part. 
\begin{table}[!htb]
  \centering
  \caption{Load-haul-dump data. Failure times in hours. The data
    are time censored at 2000 hours. }
    \vspace{3mm}
    
  \begin{tabular}[!htb]{rrrrrrrrrrrr}
  16&   39&   71&   95&   98&  110&  114&  226&  294&  344&  555&  599\\ 
 757&  822&  963& 1077& 1167& 1202& 1257& 1317& 1345& 1372& 1402& 1536\\ 
1625& 1643& 1675& 1726& 1736& 1772& 1796& 1799& 1814& 1868& 1894&
1970\\     
  \end{tabular}
  \label{tab:LHDdata}
\end{table}

\begin{table}[!htb]
  \centering
  \caption{Load-haul-dump data. Parameter estimates using methods of Section~\ref{subsec:paramest}}. 
    
    \vspace{3mm}
    
    \vekk{ 
  \begin{tabular}[!htb]{lccc}
  Estimators & $\mu$ & $\sigma$ & $\cv$  \\ \hline
  $\hat \mu, \hat \sigma, \hat \cv$ & 54.72 & 48.61 & 0.888 \\
  $\tilde \mu, \tilde \sigma, \tilde \cv$ & 55.56 & 47.23 & 0.850 \\
  Parametric: Weibull   & 55.46 & 47.22 & 0.851 \\
  \hline   
  \end{tabular}
    } 
  
  \begin{tabular}[!htb]{lccc}
  Estimators & $\mu$ & $\sigma$ & $\cv$  \\ \hline
  Sample estimators - not including censored time  & 54.72 & 48.61 & 0.888 \\
  Sample estimators - including censored time & 55.56 & 47.23 & 0.850 \\
  Parametric: Weibull - including censored time  & 55.46 & 47.22 & 0.851 \\
  \hline   
  \end{tabular}

  \label{esttab}
\end{table}

For estimation of the coefficient of variation under the null hypothesis, we estimated the parameters $\mu,\sigma,\cv$
using methods considered in Section~\ref{subsec:paramest}. The results are given in
Table~\ref{esttab}. It is seen that the  estimates which use the censored time are very close, while the one that disregard this time gives a slightly higher estimated coefficient of variation. This might be a
coincidence, however, and will not be generally valid.

In order to calculate the LR-test statistic (\ref{onestar}), we first
calculated the Laplace test statistic, and then
divided by the estimated coefficient of variation, to get $0.605/0.888 =
0.681$ using the estimates in the first row of Table~\ref{esttab}. This gave the
$p$-value 
0.50 for a two-sided test. We also
calculated the estimator $\sigma^*$ of (\ref{varest}), which gave the result
42.77, which is lower than the estimates of $\sigma$ in Table~\ref{esttab}, and would give an estimated coefficient of variation of $42.77/54.72 = 0.782$ and a test statistic of $0.605/0.782 = 0.774 $ and a $p$-value of 0.44. This illustrates the effect of using $\sigma^*$, as
estimator of $\sigma$, as discussed in Section~\ref{subsec:paramest}, namely to possibly give a lower 
estimated coefficient of variation, and in turn a lower calculated $p$-value.


Two-sided $p$-values for all tests are reported in
Table~\ref{tab:LHDdatatestresults}. In the extended Lewis-Robinson
test we used $a=1/2$, and it is interesting to see that this test
detected a significant trend in the data while the tests for
monotonic trend had fairly high $p$-values. The example thus illustrates the need for trend
tests with power against non-monotonic trend. 

\begin{table}[!htb]
  \centering
  \caption{Load-haul-dump data. The table reports $p$-values for trend tests applied to the
    load-haul-dump machine data in
    Table~\ref{tab:LHDdata}. Abbreviations: LR = Lewis-Robinson, KS =
    Kolmogorov-Smirnov, CvM = Cram$\acute{\mbox{e}}$r-von Mises, AD =
    Anderson-Darling,  ELR = Extended Lewis-Robinson test. In ELR
    $a=1/2$ was used.}
    
    \vspace{3mm}
        
  \begin{tabular}[!htb]{ccccccc}
  LR &  KS &  CvM&    AD&     ELR \\ \hline
 0.50&  0.29 &0.13& 0.086 &   0.011 \\     
  \end{tabular}
  \label{tab:LHDdatatestresults}
\end{table}

\subsection{Small Bowel Motility Data (Aalen and Husebye, 1991)}
\label{aalenhusebye}

\citeN{aalenhusebye} studied data on small bowel motility measured
on 19 persons. In particular they considered data on the length of a
cyclic motility pattern observed during a fasting state.  The data
are time censored, and each person had from one to nine complete
cycles observed before the censoring, see \citeN{aalenhusebye} for the
complete data set. 

Since the number of periods for each patient are small, and our methods are constructed for the case when censoring times $\tau$ and number of events $N(\tau)$ tend to $\infty$, we will consider testing of the null hypothesis that the 19 processes are independent RPs with the same distribution of interevent times. We therefore estimate common parameters $\mu,\sigma,\cv$ using all the data. 

It should be noted here that \citeN{aalenhusebye} fitted a model
where the events for each patient follow a Weibull RP, with individual
variation modeled by a gamma frailty model. The variation was,
however, not found significant ($p$-value 0.11), and this justifies to
some extent our analysis. On the other hand, \citeN{aalenhusebye} did
not check the data for a trend, which is the purpose of the present
example.

\begin{figure}[!hbt]
\begin{center}
\includegraphics[width=0.55\linewidth]{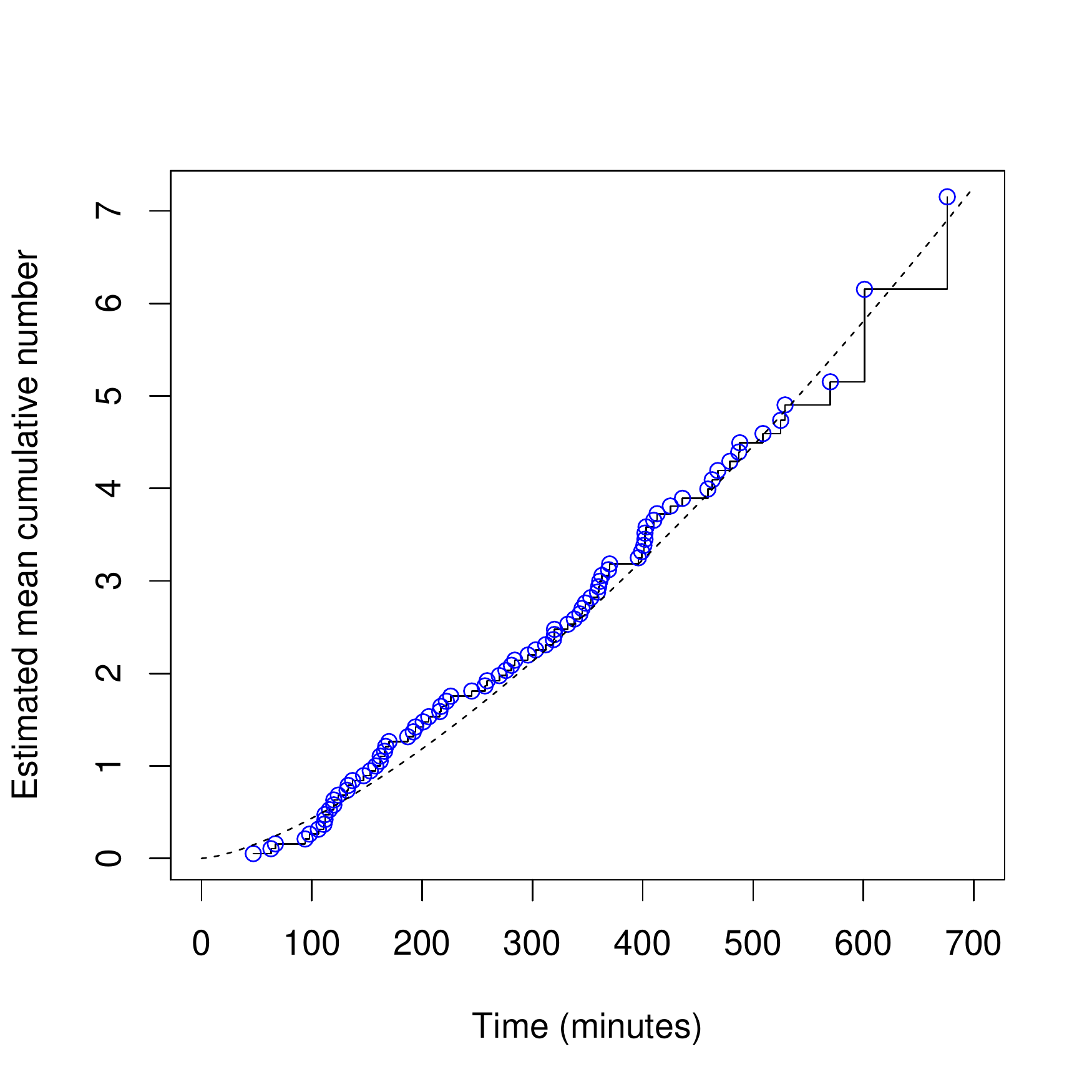}
\end{center}
\caption{Small bowel motility data. Plot of the the cumulative number of failures over time for
  the small bowel motility data. A parametric curve of the form $a t^b$ has been fitted to the data.}
\label{mtbfig}
\end{figure}

Figure~\ref{mtbfig} shows the Nelson-Aalen estimate of the common mean function $E[N(t)]$ for the patients, see \citeN{nelson88} and \citeN{lawlessnadeau} for the motivation and validity of the plot. As shown by \citeN{lawlessnadeau}, the Nelson-Aalen estimator is unbiased and consistent  for $E[N(t)]$ under fairly general conditions. Here we present the plot as an illustration of an apparent increasing trend in the data. 

In order to calculate  test statistics we need an estimate for the coefficient of variation. By considering only the 80 fully observed periods we got $\hat \mu = 98.76$, $\hat \sigma = 52.62$ and from this $\hat \cv = 52.62/98.76 = 0.533$. Since there are 19 censored interevent times in these data, one for each patient, we found
that the estimators $\tilde \mu$ and $\tilde \sigma$ are less
satisfactory. Instead we fitted a Weibull RP to the data, taking
into account the censored periods. The resulting estimates were
$\check{\mu} = 104.49, \check{\sigma}=52.45$ and $\check{\cv} =
0.502$.
There is thus a clear underdispersion in the interevent times when
comparing to the exponential distribution. Using $\hat \cv$, the
LR-statistic (\ref{lrm}) equals $1.95/0.533 = 3.67$, where 1.95 is the
value of the corresponding Laplace test statistic. Thus the $p$-value
would be  0.051 for testing the null hypothesis of HPP versus a
monotonic trend, while it is 0.00024 for the LR-test for the null
hypothesis of RP. The $p$-values obtained for different tests are
reported in Table~\ref{tab:SBMdataresults}. We see that all the tests
find a significant trend in these data.
\vekk{ 
Still the data are the same, depicted by the curve
Figure~\ref{mtbfig}. An intuitive explanation of why we would be
``more sure'' of an increasing trend from this curve if we know there
is underdispersion, is that the variability of interevent times is
then larger for a Poisson model, implying a higher risk that the
curvature was due to chance.  
} 

We also performed a parametric trend test using the TRP with a Weibull renewal distribution and a power law trend function, see Section~\ref{sec:sim}. Leaving out further details, we report a $p$-value for trend of 0.041 using a standard asymptotic likelihood ratio test.

\begin{table}[!htb]
  \centering
  \caption{Small bowel motility data. The table report $p$-values for trend tests applied to
    the data. Abbreviations: LR = Lewis-Robinson, CvM =  Cram$\acute{\mbox{e}}$r-von Mises,  AD = Anderson-Darling,  GL = Generalized Laplace Test.}
    
    \vspace{3mm}
    
  \begin{tabular}[!htb]{ccccccc}
       LR &    CvM     &    AD      &       GL \\ \hline
 $<0.0001$ & $<0.0001$& $<0.0001$&        0.007    
  \end{tabular}
  \label{tab:SBMdataresults}
\end{table}

\section{Conclusion}
\label{sec:conc}

We have presented a novel class of tests for trend in time censored
recurrent event processes, based on the general null hypothesis of an
RP. This class includes, among other tests, new versions of the
Lewis-Robinson test and the Anderson-Darling test, extending these
tests to time censored processes. For the single process case, the
Anderson-Darling test turns out to have attractive properties when
used as a test for general alternatives, both monotonic and
non-monotonic trends. If power against monotonic trends is of main
interest, the Lewis-Robinson type test is on the other hand a safe
choice, both for single and multiple processes.

The derived test statistics are based on asymptotic results for
renewal processes. The calculated critical values are hence only
approximate when used in small and medium sized samples. The
simulation study shows, however, satisfactory performance of the
tests, with some exceptions in cases with very small sample. In such
cases an alternative procedure would be to simulate 
the null distribution of the test statistic by a permutation approach,
permuting the order of the completely observed interevent times.
\citeN{Lawless2012tmt} showed that this is a valid approach even for
time censored processes, and we have confirmed this in simulations not
reported here.

It is clear that the basic result of Corollary~\ref{cor1} in principle
may give rise to a very large class of tests. We have in
Section~\ref{sec:classof}  considered four tests based on  standard
goodness-of-fit statistics, and as an example of the variety of other
possible tests we added and studied in some detail a non-standard
test,  which led to a further
extension of the Lewis-Robinson test.  

An interesting fact of the constructed test statistics is that they
may be viewed as test statistics for the case of Poisson processes,
with null hypothesis corresponding to HPP, that are adjusted according
to the coefficient of variation of the observed interevent times. This
is exactly the way \citeN{lewisrobinson} obtained their test statistic
for the event censored case, starting from the Laplace test. 

R-code for the tests can be obtained from the authors.

\vekk{ 
AVSNITTET NEDENFOR KAN MULIGENS TILPASSES FOR BRUK I CONCLUSIONS:
In practice there may as well be censoring schemes that combine the
two types. \citeN{aalenhusebye} discuss these schemes and other
schemes and point to the importance of considering censoring times
that are formally \textit{stopping times}.  
} 

\bigskip

\section*{Appendix 1}

\subsection*{Consistent Estimator of $\mu$}
It is clear from the strong law of large numbers for renewal processes (see, e.g., \citeN{ross}) that
\[
    \hat \mu = \frac{ \sum_{i=1}^{N(\tau)} X_i}{N(\tau)} \rightarrow \mu \mbox{ a.s.}
    \]
    since $N(\tau) \rightarrow \infty$. 
Note that by standard renewal process theory  we have
\[
\frac{N(\tau)}{\tau} \rightarrow  \frac{1}{\mu} \mbox{ a.s.}
\]
Thus another consistent estimator of $\mu$ is given by $\tilde \mu=\tau/N(\tau)$. Note that we can write $\hat \mu = T_{N(\tau)}/N(\tau)$, so we have $\tilde \mu > \hat \mu$.

\subsection*{Consistent Estimator of $\sigma^2$}
 
 By the strong law of large numbers we have
\[
   \frac{1}{N(\tau)} \sum_{i=1}^{N(\tau)} (X_i-\mu)^2 \rightarrow  \sigma^2 \mbox{ a.s.}
\]
Writing 
\[
   \frac{1}{N(\tau)} \sum_{i=1}^{N(\tau)} (X_i-\mu)^2
   = \frac{1}{N(\tau)} \sum_{i=1}^{N(\tau)} (X_i-\hat \mu)^2 + (\hat \mu - \mu)^2 
 \]
 it follows from Slutsky's theorem that
 \[
    \hat \sigma^2 = \frac{1}{N(\tau)} \sum_{i=1}^{N(\tau)} (X_i-\hat \mu)^2
 \]
 is a consistent estimator of $\sigma^2$. 
 
A disadvantage of the estimator $\hat \sigma$, as with $\hat \mu$, is that they do not take into account the censored time $\Y$.  \citeN[chap. 5]{gallager} shows that
 \bq
 \label{gall}
     \lim_{\tau \rightarrow \infty} \frac{\int_0^\tau (t-T_{N(t)})dt}{\tau}
     = \frac{\E(X^2)}{2\mu} \; \mbox{ a.s.}
     \eq
Here the left hand side is the long run average length of time since
the last previous event, and the result says that this equals
$(\mu/2)(1+\cv^2)$ where $\cv$  is the coefficient of
variation of the distribution of $X$. 

We use (\ref{gall}) in the following way. A straightforward calculation shows that
\[
\int_0^\tau (t-T_{N(t)})dt = \frac{1}{2} \left\{ \sum_{i=1}^{N(\tau)} X_i^2 +
(\tau - T_{N(\tau)})^2 \right\},
\]
which after substitution in (\ref{gall}), noting that $\tau/N(\tau) \rightarrow \mu$, gives  the following consistent estimator for $\sigma^2$,
    \[
    \tilde \sigma^2 = 
    \frac{1}{N(\tau)} \left\{ \sum_{i=1}^{N(\tau)} X_i^2 + (\tau - T_{N(\tau)})^2 \right\} - \tilde \mu^2.
    \]
    
\bigskip

An alternative variance estimator, $\sigma^{*2}$,  was presented in Section~\ref{subsec:paramest}, see equation (\ref{varest}).
To prove consistency of $\sigma^{*2}$ under the null hypothesis of RP,  we can consider separately the sum over odd $i$ and even $i$ and use the strong law of large numbers on each of the two resulting sums, which are now sums of i.i.d. variables.

\section*{Appendix 2}

\subsection*{The Extended Lewis-Robinson Test}
The test statistic (\ref{vvtest}) is obtained as follows.
Note first that we can write
\[
  N(t) = \sum_{i=1}^{N(\tau)} I_{[T_i,\tau)}(t),
\]
where $I_A(t)$ is the indicator function of the set $A$. 
 From (\ref{tildeeq}) it follows that we can consider  the integration
\begin{eqnarray}
\tau \int_0^{a} (N(s\tau)-sN(\tau))ds &=&
\int_0^{\tau a} \left(N(t)- \frac{N(\tau)}{\tau}t\right)dt \nonumber \\
&=&
 \left(\sum_{i=1}^{N(\tau)}\int_0^{a \tau} I_{[T_i,\tau)}(t)dt- \int_0^{a \tau} \frac{N(\tau)}{\tau}t\right)dt  \nonumber \\
&=&
\sum_{i=1}^{N(\tau)} \left(a \tau - \min \left\{T_i,a \tau \right\}\right) - 
\frac{1}{2} a^2 \tau N(\tau)
\label{vv1}
\end{eqnarray}
and similarly
\begin{eqnarray}
\tau \int_{a}^1 (N(s\tau)-sN(\tau))ds &=&
\int_{ a\tau}^\tau \left(N(t)- \frac{N(\tau)}{\tau}t\right)dt \nonumber \\
&=&
 \left(\sum_{i=1}^{N(\tau)}\int_{a \tau}^\tau I_{[T_i,\tau)}(t)dt- \int_{a \tau}^\tau \frac{N(\tau)}{\tau}t\right)dt \nonumber \\
&=&
\sum_{i=1}^{N(\tau)} \left(\tau - \max \left\{T_i, a \tau\right\}  \right) - \frac{1}{2}  (1-a^2) \tau  N(\tau). \label{vv2}
\end{eqnarray}
Subtracting (\ref{vv2}) from (\ref{vv1}), we get
\[
 \sum_{i=1}^{N(\tau)} \left[  \left(a \tau - \min \left\{T_i,a \tau\right\}\right) -  \left(\tau - \max \left\{T_i,a \tau\right\}  \right) \right]  + \left( \frac{1}{2} - a^2 \right) \tau N(\tau)
\]
\[
    =  \sum_{i=1}^{N(\tau)}   |T_i - a \tau| 
    - \left( \frac{1}{2} - a(1-a) \right) \tau N(\tau)
\]
The statistic (\ref{vvtest}) is hence obtained from (\ref{tildeeq}). 

We finally prove that $\int_0^{a} W^0(s)ds -      \int_{a}^1 W^0(s)ds$ is normal with mean 0 and variance $(1/12)-a^2(1-a)^2$. For this we use the fact that,
for a Gaussian process $G(t)$ with mean function $\E(G(t))=m(t)$ and covariance function $Cov(G(s),G(t))=k(s,t)$, we have
\[
   \int_a^b G(t)dt \sim N(\int_a^b m(t)dt, \int_a^b \int_a^b k(s,t) ds dt).
   \]
The covariance function of the Brownian bridge is $k(s,t)=\min(s,t)-st$. Hence 
   $U =\int_0^{a} W^0(s)ds$ is normal with mean 0 and variance
   \begin{equation}
   \label{varu}
      \int_0^{a} \int_0^{a} (\min(s,t)-st) ds dt = \frac{a^3(4-3a)}{12}.
   \end{equation} 
   Similarly, $V =\int_a^{1} W^0(s)ds$ is normal with mean 0 and variance
   \begin{equation}
   \label{varv}
      \int_a^{1} \int_a^{1} (\min(s,t)-st) ds dt = (1/12) - \frac{a^2(6-8a+3a^2)}{12}.
   \end{equation} 
   
Now we want $\Var(U-V) = \Var(U) + \Var(V) - 2\Cov(U,V)$, and thus we seemingly need also $\Cov(U,V)$. We use, however, the following trick. Since $U+V = \int_0^1 W^0(s)ds$ is known to have variance $1/12$, we can solve for $\Cov(U,V)$ in the equation $\Var(U+V) = \Var(U)+\Var(V)+2\Cov(U,V)$
   to get 
   \[
  \Var(U-V) = 2(\Var(U) + \Var(V)) - \Var(U+V) = \frac{1}{12} - a^2(1-a)^2 
   \]
where we used (\ref{varu}-\ref{varv}) and the fact that  $\Var(U+V) = 1/12$.

\bibliography{trendbib}

\bibliographystyle{dcu}

\end{document}